\documentclass[pre,twocolumn,twoside,showpacs,byrevtex,superscriptaddress]{revtex4-1}

\usepackage{currfile}
\usepackage{hyperref}
\usepackage{graphicx,epsfig,verbatim,enumerate}
\usepackage{amssymb,amsmath,mathdots}
\usepackage{ifthen}
\usepackage{rotating}
\usepackage{sidecap}
\usepackage{longtable}
\usepackage{afterpage}

\usepackage{enumitem}

\usepackage{array}
\usepackage{amsmath}
\usepackage{amsfonts}
\usepackage{algorithm}
\usepackage[noend]{algpseudocode}

\makeatletter
\def\BState{\State\hskip-\ALG@thistlm}
\makeatother

\newboolean{twocolswitch}

\newcommand{\tavg}[1]{\langle#1\rangle}

\newcommand{\sindex}[1]{}
\newcommand{\nindex}[1]{}

\newcommand{\etal}{\textit{et al.}}
\newcommand{\www}[1]{\url{#1}}

\newcommand{\Req}[1]{Eq.~(\ref{#1})}

\usepackage{lettrine}

\usepackage{xcolor}

\definecolor{verylightgray}{rgb}{0.9,0.9,0.9}
\definecolor{grey}{rgb}{0.5,0.5,0.5}

\newcommand{\calin}{C_{\textrm{in}}}
\newcommand{\calout}{C_{\textrm{out}}}
\newcommand{\calrat}{C_{\textrm{rat}}}
\newcommand{\caldiff}{C_{\textrm{diff}}}

\newcommand{\calinfn}[1]{C_{\textrm{in}}(#1)}
\newcommand{\caloutfn}[1]{C_{\textrm{out}}(#1)}
\newcommand{\calratfn}[1]{C_{\textrm{rat}}(#1)}

\newcommand{\calinoutfnsup}[1]{C_{\textrm{i/o}}^{#1}}
\newcommand{\calinoutfn}[1]{C_{\textrm{i/o}}(#1)}
\newcommand{\frequency}[2]{f(#1|\,#2)}
\newcommand{\normfrequency}[2]{p(#1|\,#2)}

\newcommand{\textsymbol}{T}
\newcommand{\phrasesymbol}{s}
\newcommand{\phrasesetsymbol}{S}

\newcommand{\textcomp}{\textsymbol_{\textrm{comp}}}
\newcommand{\textref}{\textsymbol_{\textrm{ref}}}

\newcommand{\foodphrase}{\phrasesymbol}
\newcommand{\foodphraseset}{\phrasesetsymbol_{\textrm{in}}}
\newcommand{\activityphrase}{\phrasesymbol}
\newcommand{\activityphraseset}{\phrasesetsymbol_{\textrm{out}}}

\newcommand{\foodactivityphrase}{\phrasesymbol}
\newcommand{\foodactivityphraseset}{\phrasesetsymbol_{\textrm{i/o}}}

\newcommand{\rhopearson}{\hat{\rho}_{\textrm{p}}}
\newcommand{\rhospearman}{\hat{\rho}_{\textrm{s}}}

\setboolean{twocolswitch}{true}

\newcommand{\revtexonly}[1]{#1}
\newcommand{\plainlatexonly}[1]{}
\newcommand{\revtexlatexswitch}[2]{#1}
\newcommand{\plosoneonly}[1]{}

\begin{document}

\title{\protect
The Lexicocalorimeter:  \\
Gauging public health through caloric input and output on social media
}

\author{
  \firstname{Sharon E.}
  \surname{Alajajian}
}

\affiliation{
  Department of Mathematics \& Statistics,
  Vermont Complex Systems Center,
  Computational Story Lab,
  \& the Vermont Advanced Computing Core,
  The University of Vermont,
  Burlington, VT 05401.
}

\author{
  \firstname{Jake Ryland}
  \surname{Williams}
}

\affiliation{
  School of Information,
  University of California Berkeley,
  102 South Hall \#4600,
  Berkeley, CA 94720-4600.
}

\author{
  \firstname{Andrew J.}
  \surname{Reagan}
}

\affiliation{
  Department of Mathematics \& Statistics,
  Vermont Complex Systems Center,
  Computational Story Lab,
  \& the Vermont Advanced Computing Core,
  The University of Vermont,
  Burlington, VT 05401.
}

\author{
  \firstname{Stephen C.}
  \surname{Alajajian}
}

\affiliation{
  Women, Infants, and Children,
  East Boston, MA 02128.
}

\author{
  \firstname{Morgan R.}
  \surname{Frank}
}

\affiliation{
  Media Lab,
  Massachusetts Institute of Technology,
  Cambridge,
  MA, 02139
}

\author{
  \firstname{Lewis}
  \surname{Mitchell}
}

\affiliation{
  School of Mathematical Sciences,
  North Terrace Campus,
  The University of Adelaide,
  SA 5005, Australia
}

\author{
  \firstname{Jacob}
  \surname{Lahne}
}

\affiliation{
  Culinary Arts and Food Science,
  Drexel University,
  3141 Chestnut Street, 
  Philadelphia, PA 19104.
}

\author{
  \firstname{Christopher M.}
  \surname{Danforth}
}

\affiliation{
  Department of Mathematics \& Statistics,
  Vermont Complex Systems Center,
  Computational Story Lab,
  \& the Vermont Advanced Computing Core,
  The University of Vermont,
  Burlington, VT 05401.
}

\author{
  \firstname{Peter Sheridan}
  \surname{Dodds}
}

\email{salajajian@asc.upenn.edu,
\hfill
lewis.mitchell@adelaide.edu.au,\\
jake.williams@drexel.edu,
\hfill
jl3542@drexel.edu,\\
andrew.reagan@uvm.edu,
\hfill
chris.danforth@uvm.edu,\\
stephenalajajian@gmail.com,
\hfill
peter.dodds@uvm.edu.\\
mrfrank@mit.edu,
}

\affiliation{
  Department of Mathematics \& Statistics,
  Vermont Complex Systems Center,
  Computational Story Lab,
  \& the Vermont Advanced Computing Core,
  The University of Vermont,
  Burlington, VT 05401.
}

\date{\today}

\begin{abstract}
  \protect
  We propose and develop a Lexicocalorimeter: 
an online, interactive instrument for measuring the ``caloric
content'' of social media and other large-scale texts.
We do so by constructing extensive yet improvable
tables of food and activity related phrases,
and respectively assigning them with sourced estimates of caloric 
intake and expenditure.
We show that for Twitter, our naive measures of 
``caloric input'', ``caloric output'', and
the ratio of these measures are all strong
correlates with health and well-being measures for the
contiguous United States.
Our caloric balance measure in many cases outperforms both its constituent quantities;
is tunable to specific health and well-being measures such as diabetes rates;
has the capability of providing 
a real-time signal reflecting a population's health;
and has the potential to be used alongside
traditional survey data in the development 
of public policy and collective self-awareness.
Because our Lexicocalorimeter is a linear superposition
of principled phrase scores,
we also show we can move beyond correlations to explore what people talk about in 
collective detail, and
assist in the understanding and explanation of how population-scale conditions vary,
a capacity unavailable to black-box type methods.

\end{abstract}

\maketitle

\section{Introduction}
\label{sec:fluxwell.intro}

Online instruments designed to measure social,
psychological, and physical well-being at a population level are 
becoming essential for public policy purposes and 
public health monitoring~\cite{CDC2013,dodds2011e}.
These data-centric gauges both 
empower the general public with information
to allow comparisons of communities at all scales,
and naturally
complement the broad, established
set of more readily measurable socioeconomic indicators 
such as wage growth, crime rates, and housing prices.

Overall well-being, or quality of life, depends on  many factors 
and is complex to measure~\cite{diener1995}.
Existing techniques for estimating population well-being range from traditional surveys~\cite{CDC2013,gallup}
to estimates of smile-to-frown ratios captured automatically
on camera in public spaces~\cite{feelometer}, and vary widely in the types of data they amass,
collection methods, cost, time scales involved, and
degree of intrusion.
Partly in response to policy makers' desire 
for simple ``one number''
quantification of complex systems---arguably a general
human proclivity---many measures are composite in nature.
Two examples are
(1) the Gallup Well-Being Index, 
which is based on factors such as life evaluation, emotional health, physical health,
healthy behavior, work environment, and basic access to necessary
resources~\cite{gallup};
and
(2) the Living Conditions measure developed by the United States Census Bureau,
which is derived from housing conditions,
neighborhood conditions, basic needs met, a ``full set'' of
appliances, and access to help if needed~\cite{census}.

While such measures will always have their place,
we venture that we must resist oversimplification.
The dashboard of society should be just that---a rich set
of incompatible instruments whose informational content may be observed individually
and in total, not unlike the required input needed for flying a plane
where knowledge of just a single number representing ``things are
going well'' would be untenable.
The construction of data-centric instruments for social systems
that deliver more direct, interpretable measures is 
therefore of great importance as we move forward into 
the age of ubiquitous (but not complete) measurement.

With the explosive growth of online activity and social
media around the world, the massive amount of real-time
data created directly by populations
of interest has become an increasingly attractive and fruitful
source for analysis.
Despite the
limitation that social media users in the United States 
are not a random sample of the US 
population~\cite{pew}, 
there is a wealth of information in these data sets and uneven
sampling can often be accommodated.

Indeed, online activity is now considered by many to be a promising 
data source for detecting health
conditions~\cite{signorini2011,prieto2014} and 
gathering 
public-health
information~\cite{chew2010a,paul2011},
and within the last decade, researchers have constructed a range of
online public-health instruments with varying degrees of success.
The maturing of these and related instruments along with
theoretical models will 
ultimately fundamentally inform the limits of characterization 
and predictability of social systems.

In the next two subsections, 
we cover related research and then 
describe our approach 
to measuring the ``caloric content'' of text.

\subsection{Previous work}
\label{subsec:fluxwell.previouswork}

For a general overview of work relevant to our present effort,
we briefly summarize related research concerning public health and well-being
in connection with a range of social media and online data sets.

In the difficult realm of predicting pandemics~\cite{watts2005a}, 
Google Flu Trends~\cite{googleflutrends2015a} enjoyed early success
and acclaim.
Initially based very simply on search terms,
the instrument proved unsurprisingly
to be imperfect and in need of a more 
sophisticated approach~\cite{lazer2014a}.

In work by several of the current authors and colleagues,
Mitchell \etal\ measured the happiness of tweets across the US  and found
strong correlations with other indices of well-being at city
and state level, such as the
Gallup Well-being Index; the Peace Index; the America's Health Ranking
composite index of Behavior, Community and Environment, Policy and
Clinical Care metrics; and gun violence (negative
correlation)~\cite{mitchell2013a}. 
Using the same instrument in 10 languages, the Hedonometer, 
we have also shown that 
the emotional content of tweets tracks major world
events~\cite{dodds2011e,dodds2015a}.

Paul and Dredze found that states with higher
obesity rates have more tweets about obesity, and states with higher
smoking rates have more tweets about cancer~\cite{paul2011}.  They also found a
negative correlation between exercise and frequency of tweeting about
ailments, suggesting ``Twitter
users are less likely to become sick in states where people exercise.''
They further found health care coverage rates to be
negatively correlated with likelihood of posting tweets about
diseases.  

Chunara \etal\ recently found that
activity-related interests on Facebook are negatively correlated with
being overweight and obese, while interest in television is positively
correlated with the same~\cite{chunara2013}.  

In an analysis of online recipe queries, West \etal\ found
that the number of patients admitted to the emergency room of a major
urban hospital in Washington, DC for congestive heart failure (CHF)
each month was significantly correlated with average sodium per recipe
searched for on the Web in the same month~\cite{west2013}.

Eichstaedt and colleagues~\cite{eichstaedt2015a} have demonstrated
that psychological language on Twitter outperforms certain composite 
socioeconomic indices in predicting heart disease at the county level.
They were able to show in particular that the expression
of negative emotions such as
anger on Twitter could be taken as a kind of risk factor at
the population scale.

On a US county level, Culotta~\cite{culotta2014a}
found that Twitter activity
provided a more ``fine-grained representation'' of
community health than demographics alone with the
prevalance of particular words that indicate, for example,
television habits, or negative engagement.

Finally, in work directly related to our present study,
Abbar \etal~\cite{abbar2014a} have recently
performed a similar analysis 
of translating food terms used on Twitter into calories.
They found a correlation between
Twitter calories and obesity and diabetes rates for the US,
and explored how food-themed interactions over social networks vary
with connectedness, finding suggestions of social contagion.
While our approach and results are largely sympathetic,
our work incorporates estimates of physical activity which we
will show provides essential extra information regarding health;
introduces a phrase extraction method we call serial partitioning;
and leads to an online
implementation, paving the way for a real-time instrument as part of our
proposed `panometer.'
We also note that we carried out our work concurrently
and independently.

\subsection{Lexicocalometrics}
\label{subsec:fluxwell.now}

From the preceding list of studies, it has become clear that we 
can estimate population-scale levels of health and well-being through social media.
Here, we examine the words and phrases people post publicly about food and physical
activity on Twitter on a statewide level for the contiguous United
States (48 states along with the District of Columbia).
As we explain fully below in
\revtexlatexswitch{
  Sec.~\ref{subsec:fluxwell.calories} and Methods and Materials, Sec.~\ref{sec:fluxwell.methods},
}{
  Estimating~Calories~from~Phrases in the Analysis~and~Results section,
  and in Methods~and~Materials,
}
we group categorically similar words and phrases into lemmas,
and we then assign caloric values to these lemmas
using the terms and notation ``caloric input'' for food, $\calin$,
and ``caloric output'' for activity, $\calout$.
We define the ratio of caloric output to caloric input
to be a third quantity, ``caloric ratio'':
\begin{equation}
  \calrat
  = 
  \frac{\calout}{\calin}.
  \label{eq:fluxwell.calrat}
\end{equation}
While we will focus largely on the three
quantities 
$\calin$,
$\calout$,
and
$\calrat$,
we will also explore 
``caloric difference'',
an alternate combination
of $\calin$ and $\calout$ involving a single parameter:
\begin{equation}
  \caldiff(\alpha)
  =
  \alpha \calout 
  - 
  (1-\alpha)\calin,
  \label{eq:fluxwell.caldiffalpha}
\end{equation}
where 
$ 0 \le \alpha \le 1$.
We use ``phrase shifts''~\cite{dodds2011e} to show how 
specific
lemmas---e.g., ``apples'',
``cake with frosting'', 
``white water rafting'',
``knitting'',
and ``watching tv or movie''
contribute to the caloric texture of states across the contiguous US.
We then correlate all three values with 37
measures relating to health and well-being, 
and we find statistically
strong correlations with quantities such as high blood pressure, inactivity,
diabetes levels, and obesity rates.
For ease of language, we will generally speak of 
phrases rather than lemmas.

We have also generated an accompanying online, interactive
instrument for exploring health patterns through the lens of ``Twitter
calories'': the Lexicocalorimeter.
An initial, fixed version of the instrument may be accessed at this paper's
Online Appendices,
\url{http://compstorylab.org/share/papers/alajajian2015a/},
with a evolvable, production version housed within our larger
measurement platform \url{http://panometer.org}
at 
\url{http://panometer.org/instruments/lexicocalorimeter}
(all code for these sites can be found at
\url{https://github.com/andyreagan/lexicocalorimeter-appendix}).
We note that while our online instrument is based on Twitter, it may in principle
be used on any sufficiently large text source, social media or
otherwise, such as Facebook.

From this point, we structure the core of our paper as follows.
In 
\revtexlatexswitch{Sec.~\ref{sec:fluxwell.results}, }{Sec.~Analysis~and~Results, }
we establish and discuss our findings in depth.
Specifically, we:
(1) Outline our text analysis of a Twitter corpus from 2011--2012
\revtexlatexswitch{Sec.~\ref{subsec:fluxwell.calories}),}{(see Estimating~Calories~from~Phrases in the Analysis~and~Results section),}
reserving full details for Methods and Materials in 
\revtexlatexswitch{Sec.~\ref{sec:fluxwell.methods}; }{Sec.~Methods~and~Materials; }
(2) Present caloric maps of the contiguous US
contrasting the 48 states and DC
through histograms and phrase shifts
\revtexlatexswitch{(Sec.~\ref{subsec:fluxwell.maps}); }{(see Caloric~Maps~of~the~Contiguous~US in Methods~and~Materials); }
and 
(3) Examine how  $\calin$, $\calin$, $\calrat$, and $\caldiff(\alpha)$
correlate with a suite of measures relating to health and well-being.
In the Supporting Information, 
we provide a sample of confirmatory figures 
as well as all shareable data sets (e.g., IDs for all tweets).
We offer concluding thoughts in
\revtexlatexswitch{Sec.~\ref{sec:fluxwell.conclusion}.}{Concluding Remarks. }

\section{Analysis and Results}
\label{sec:fluxwell.results}

\subsection{Estimating calories from phrases}
\label{subsec:fluxwell.calories}

We used all available geotagged tweets from 2011 and 2012
(around 50 million)
from a bounding box of the contiguous US, 
using Twitter's garden hose sample
(which is a sample of approximately 10\% of all tweets,
including those that are not geotagged)
and the geotag feature
to determine from which of the 48 continental states
and the District of Columbia each tweet came.
From this sample, we counted the total number
of times each food and physical activity
phrase in our database was tweeted about
in each of the 48 continental states and the District of Columbia
(see 
\revtexlatexswitch{Sec.~\ref{sec:fluxwell.methods} }{Methods~and~Materials } 
and
\revtexlatexswitch{Dataset S1}{S1 dataset}
at \url{https://dx.doi.org/10.6084/m9.figshare.4530965.v1} for all
tweet IDs).
We then used these counts to determine the average caloric input $\calin$
from food phrase tweets
and the average caloric output $\calout$
from physical activity phrase tweets as follows.

First, we equate each food phrase $\foodphrase$
with the calories per 100 grams of that food,
using the notation
$\calinfn{\foodphrase}$.
(We also explored serving sizes but the databases 
available proved far from complete.)
We then compute the caloric input
for a given text $T$ as:
\begin{equation}
  \calinfn{T}
  = 
  \frac{
    \sum_{\foodphrase \in \foodphraseset}
    \calinfn{\foodphrase}
    \frequency{\foodphrase}{T}
  }
       {
         \sum_{\foodphrase}
         \frequency{\foodphrase}{T}
       }
       = 
       \sum_{\foodphrase \in \foodphraseset}
       \calinfn{\foodphrase}
       \normfrequency{\foodphrase}{T},
       \label{eq:fluxwell.calin}
\end{equation}
where $\frequency{\foodphrase}{T}$ 
is the frequency of phrase $\foodphrase$ 
in text $T$,
$\normfrequency{\foodphrase}{T}$
is the normalized version,
and 
$\foodphraseset$
is the set of all food phrases in our database.

Second, for each tweeted physical activity phrase, 
we use an estimate of the Metabolic Equivalent of Tasks, or METs,
which we then converted to calories expended per hour,
assuming a weight of 80.7 kilograms, the average weight of a North American
adult~\cite{walpole2012a}.
Analogous to $\calinfn{T}$ above, 
we then have
\begin{equation}
  \caloutfn{T}
  = 
  \sum_{\activityphrase \in \activityphraseset}
  \caloutfn{\activityphrase}
  \normfrequency{\activityphrase}{T},
  \label{eq:fluxwell.calout}
\end{equation}
where now 
$\activityphraseset$
is the set of all phrases in our activity database.

We emphasize that both our food and exercise phrase data sets 
and Twitter databases are necessarily incomplete in nature.
The values of $\calin$ and $\calout$ are thus not meaningful as
absolute numbers but rather have power for comparisons.
We also acknowledge that our equivalences are crude---e.g., each mention
of a specific food is naively turned into the calories associated with 100 grams of
that food---and later on we address our choices in more depth.
Nevertheless, our method is pragmatic yet---as we will
show---effective, and offers clear directions for future improvement.

For simplicity and ultimately because the results are sufficiently
strong, we did not filter tweets beyond their geographic location.
Tweets may thus come from individuals, restaurants, sports stores,
resorts, news outlets, marketers, fitness apps, tourists, and so on,
and further improvements and refinements may be achieved by
appropriately constraining the Twitter corpus.

Finally, we take the ratio of $\caloutfn{T}$ to $\calinfn{T}$
to obtain the text's caloric ratio
$\calratfn{T}$.
In general, we observe that 
a higher value of $\calratfn{T}$ at the population scale 
would appear to be intuitively better, up to some limit 
indicating negative energy balance.
We note that $\calrat = 1$ is not salient and
should not be taken to mean a population is `balanced calorically'.
As we discuss later,
using the difference, what we call Caloric Difference,
a generalization of 
$\calout 
-
\calin$,
generates similar results
but, from a framing perspective,
we have reservations in creating a scale with a 0 point 
given the approximate nature of our measures.

\subsection{Caloric maps of the contiguous US}
\label{subsec:fluxwell.maps}

\begin{figure*}[tp!]
  \begin{center}
    \includegraphics[width=0.98\textwidth]{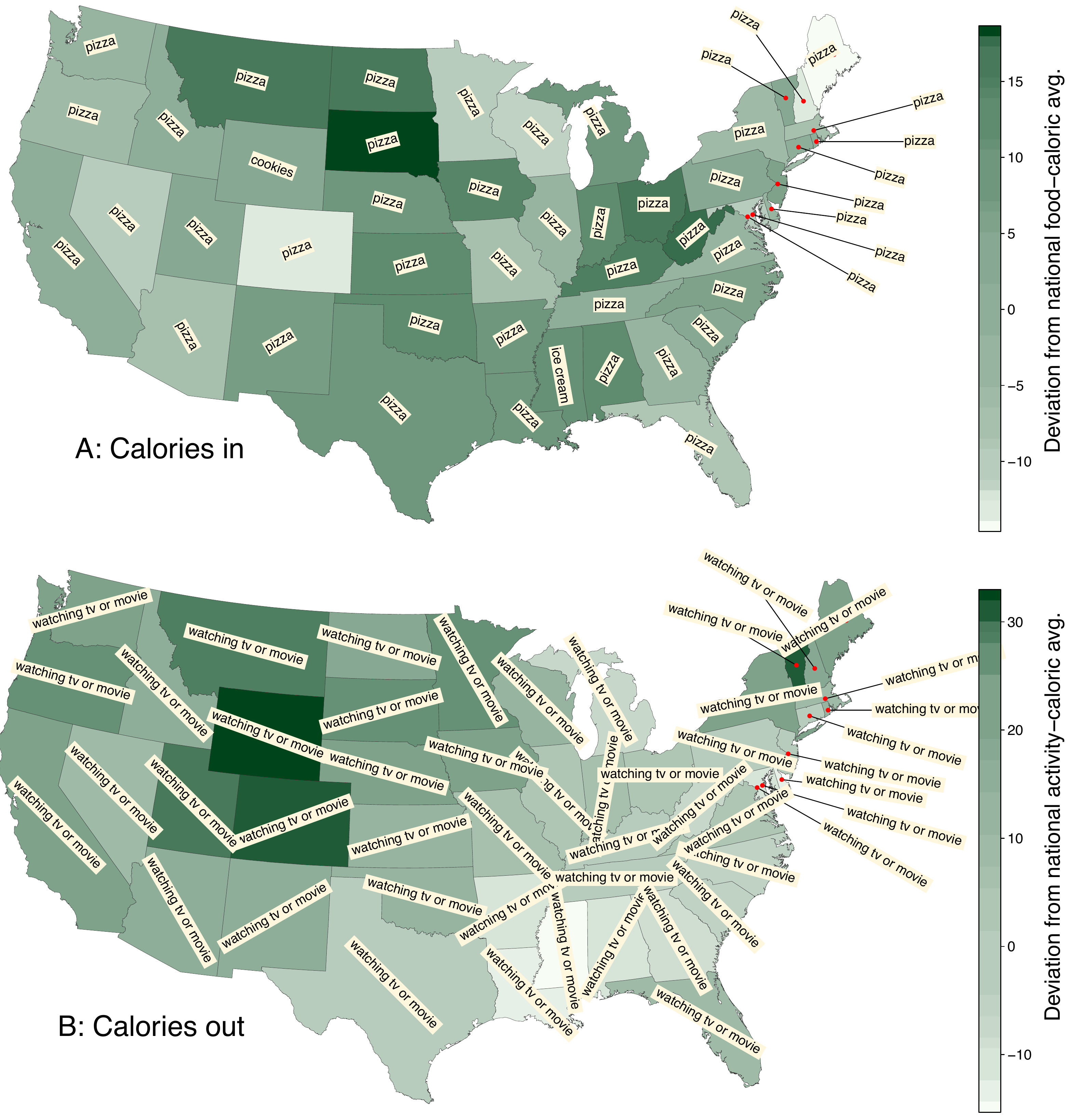}
  \end{center}
  \caption{
    Choropleth maps indicating 
    \textbf{(A)} caloric input $\calin$ and
    \textbf{(B)} caloric output $\calout$
    in the contiguous United States (including the District of 
    Columbia) based on 50 million geotagged 
    tweets taken from 2011--2012.
    For both maps, darker means higher values as per the color bars
    on the right.
    The histograms in 
    \revtexlatexswitch{Figs.~\ref{fig:fluxwell.histograms},
      \ref{fig:fluxwell.supp.histograms-food},
      and \ref{fig:fluxwell.supp.histograms-activities}}{
      Fig.~\ref{fig:fluxwell.histograms},
      \ref{fig:fluxwell.supp.histograms-food} Fig.,
      and \ref{fig:fluxwell.supp.histograms-activities} Fig.}
    show the specific rankings according to these two variables
    and also 
    $\calrat$ (see Fig.~\ref{fig:fluxwell.maps-calrat}).
    The overlaid phrase lemmas
    are the most dominant contributors to 
    $\calin$ and $\calout$---almost universally
    ``pizza'' and ``watching tv or movie''.
  }
  \label{fig:fluxwell.maps-dominantmodes}
\end{figure*}

\begin{figure*}[tp!]
  \begin{center}
    \includegraphics[width=\textwidth]{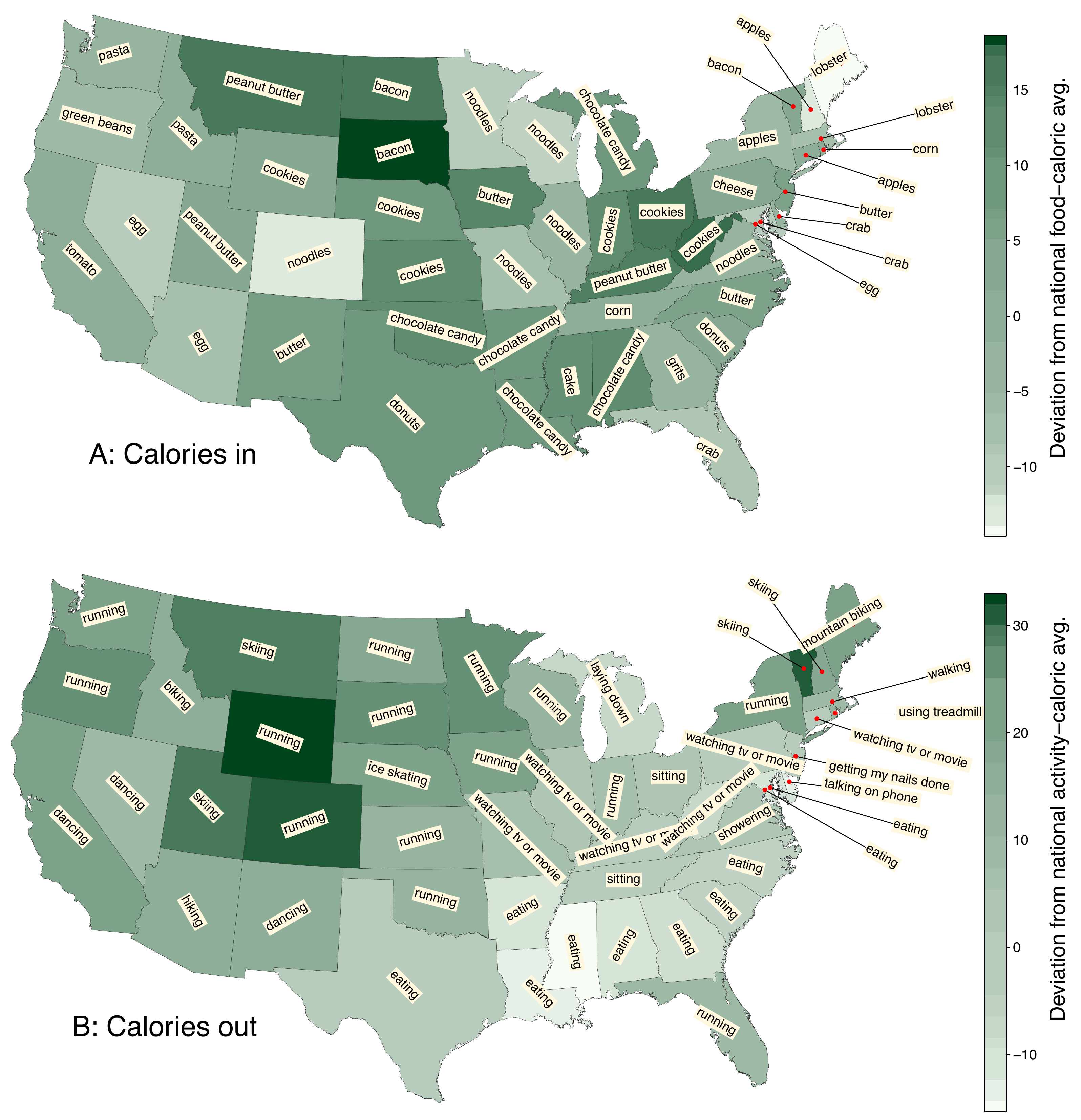}
  \end{center}
  \caption{
    The same choropleth maps for $\calin$ and $\calout$
    presented Fig.~\ref{fig:fluxwell.maps-dominantmodes}
    but now with 
    phrases whose increased usage contribute the most 
    to a population's $\calin$ and $\calout$ differing from the
    overall averages of these measures. 
    \revtexlatexswitch{See
      Sec.~\ref{subsec:fluxwell.phraseshifts}.}{See the section on Phrase~Shifts in Analysis~and~Results.}                                                                                                                                         
    For example, tweets from Vermont, which was above average
    for both $\calin$ and $\calout$ for 2011--2012,
    disproportionately contain
    ``bacon'' and ``skiing''.
    Michigan
    was above average for $\calin$ and below for $\calout$ in 2011--2012,
    and the most distinguishing phrases are
    ``chocolate candy'' and ``laying down''.
    See 
    \revtexlatexswitch{Figs.~\ref{fig:fluxwell.histograms},
      \ref{fig:fluxwell.supp.histograms-food},
      and \ref{fig:fluxwell.supp.histograms-activities}}{
      Fig.~\ref{fig:fluxwell.histograms},
      \ref{fig:fluxwell.supp.histograms-food} Fig.,
      and \ref{fig:fluxwell.supp.histograms-activities} Fig.}
    for ordered rankings.
  }
  \label{fig:fluxwell.maps-diffs}
\end{figure*}

We now move to our central analysis and exploration of 
how our lexicocalorimetric measure varies geographically.
We start with visual representations
and then continue on to more detailed comparisons.

In Fig.~\ref{fig:fluxwell.maps-dominantmodes},
we show two choropleth maps 
of our overall 2011--2012 measures
of Twitter's
caloric input $\calin$
and
caloric output $\calout$.
For both maps and those that follow, quantities increase as colors move from 
light to dark green.

These maps immediately allow for some basic observations
which we will delve into and harden up as our analysis proceeds.
For the food calories map, we see $\calin$
is generally largest in the Midwest and the south
while Colorado and Maine stand out as states with
the lowest calories.

We see a different texture in the activity calories 
map with the highest caloric output according to our measure appearing in the
three-state block of Wyoming, Colorado, and Utah, as well as Vermont.
Tweet-based caloric output drops to a low in Mississippi
and the surrounding states, while Michigan also
appears to have a low value of $\calout$.

For the food and activities maps in
Fig.~\ref{fig:fluxwell.maps-dominantmodes},
we also show the most dominant phrase for
each population's $\calin$ and $\calout$ scores.
Almost uniformly, ``pizza'' (high calorie food) and 
``watching tv or movie'' (low calorie activity) are 
the lemmas with the largest contributions,
a function of both volume and caloric scores.
Only Mississippi (``ice cream'') and Wyoming (``cookies'')
are exceptions, though ``pizza'' is still near the top for both.

In Fig.~\ref{fig:fluxwell.maps-diffs}, 
we present the same choropleth maps from
Fig.~\ref{fig:fluxwell.maps-dominantmodes},
but now with the phrase most distinguishing a population.
Specifically, we show phrases whose increased prevalence
most contributes to moving a population's Twitter calorie scores away
from the overall average for the contiguous US.
For example, if a population's $\calin$ is above average, we find the
food phrase
whose frequency coupled with its caloric content most strongly
moves the population's $\calin$ up from the average.
(We explain in full how we determine these phrases
later with phrase shifts
\revtexlatexswitch{in Sec.~\ref{subsec:fluxwell.phraseshifts}.)}{in Analysis~and~Results.)}
We now see a diverse spread of terms.
We find a number of phrases make for 
reasonable representations:
\begin{itemize}
\item 
  ``lobster'' in Maine and Massachusetts;
\item 
  ``grits'' in Georgia;
\item 
  ``skiing'' in Vermont, New Hampshire, and Utah;
\item 
  and
  ``running'' in Colorado and a number of other locations.
\end{itemize}

Prototypical unhealthy foods rise to the top in various states:
\begin{itemize}
\item 
  ``donuts'' in Texas;
\item 
  ``cake'' in Mississippi;
\item 
  ``chocolate candy'' in Louisiana;
\item 
  and ``cookies'' in Indiana.
\end{itemize}
By contrast, a few ``virtuous'' foodstuffs appear such as ``green beans''
in Oregon and ``tomato'' in California.

Our activity list also includes some rather low intensity
ones and we see:
\begin{itemize}
\item 
  ``eating'' rising to the top in Texas, the south, and a number other states;
\item 
  ``watching tv or movie'' in Pennsylvania and elsewhere;
\item 
  ``sitting'' in Tennessee;
\item 
  ``talking on the phone'' in Delaware;
\item 
  ``getting my nails done'' in New Jersey;
\item 
  and simply
  ``lying down'' in Michigan.
\end{itemize}

Now, we do not pretend that these phrases all
come from individuals diligently recording
their present meals or activities.
Apart from tweets from individuals, our database contains tweets from
companies, advertisers, resorts, and so on.
And some phrases are problematic in their 
generality of meaning, most especially ``running''
(the word ``run'' currently has the most meanings
in the Oxford English Dictionary).
Nevertheless, as we dig deeper into all the phrases
found for a particular state, 
we will continue to find commonsensical lexical patterns.

\begin{figure}[tp!]
  \begin{center}
    \includegraphics[width=0.5\textwidth]{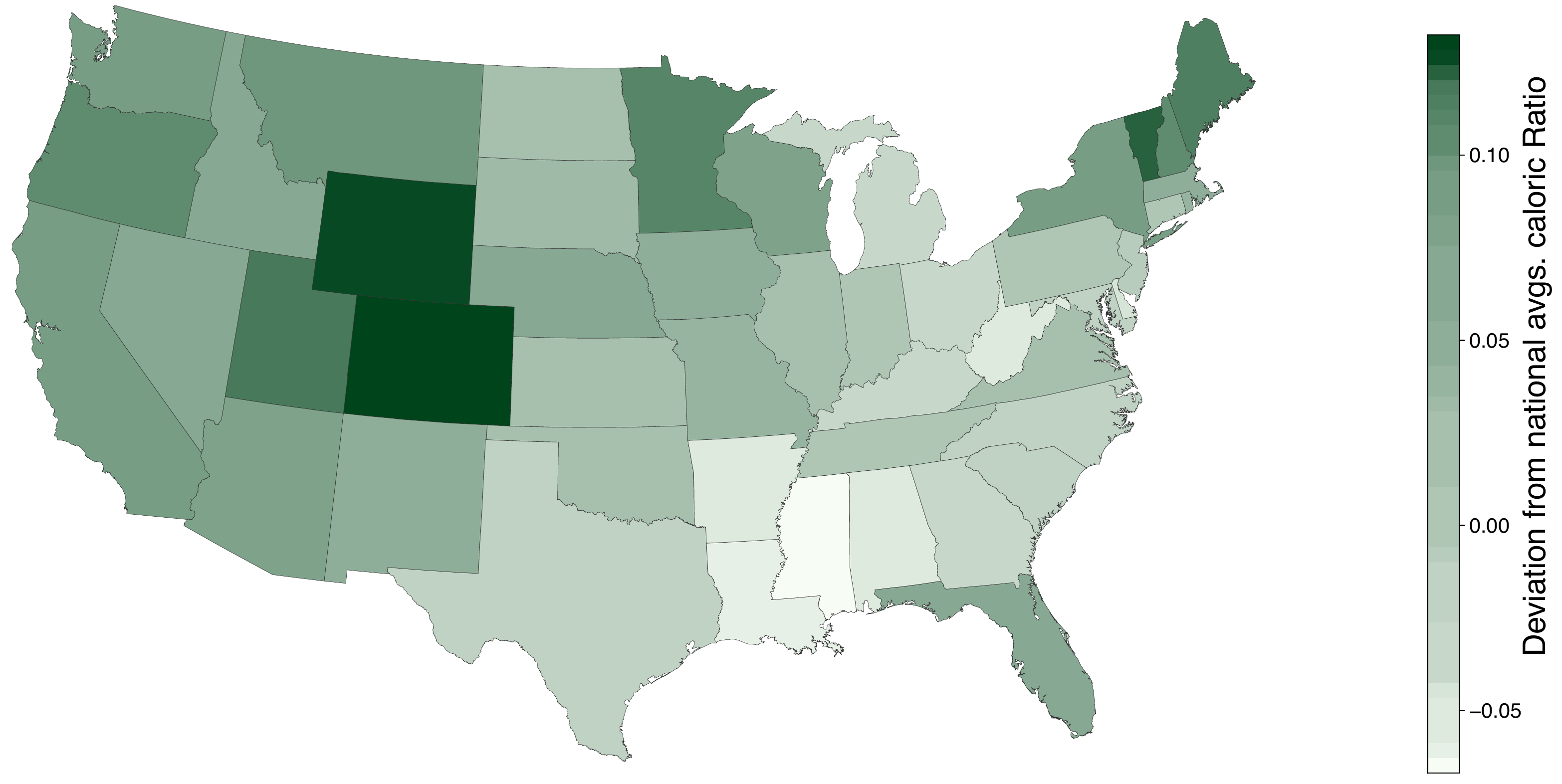}
  \end{center}
  \caption{
    Choropleth for caloric ratio $\calrat = \calout/\calin$.
    See
    \revtexlatexswitch{Figs.~\ref{fig:fluxwell.histograms},
      \ref{fig:fluxwell.supp.histograms-food},
      and \ref{fig:fluxwell.supp.histograms-activities}}{
      Fig.~\ref{fig:fluxwell.histograms},
      \ref{fig:fluxwell.supp.histograms-food} Fig.,
      and \ref{fig:fluxwell.supp.histograms-activities} Fig.}
    for ordered rankings.
  }
  \label{fig:fluxwell.maps-calrat}
\end{figure}

In Fig.~\ref{fig:fluxwell.maps-calrat},
we show a choropleth map for caloric ratio, $\calrat$.
We see that the highest values of $\calrat$ 
are found in Colorado, Wyoming, and Vermont, 
and secondarily for Maine, Minnesota, Oregon, and Utah.
Low values of $\calrat$ appear in the region 
comprising Mississippi, Louisiana, Alabama, and Arkansas,
as well as West Virginia.

An initial visual comparison of
of 
Figs.~\ref{fig:fluxwell.maps-dominantmodes}
and \ref{fig:fluxwell.maps-calrat},
suggest that $\calout$ is more well aligned
with $\calrat$ than $\calin$.
The reason is that for the present version
of the Lexicocalorimeter,
$\calout$ has a larger dynamic
range than $\calin$,
roughly 250 to 285 versus 160 to 210 
giving ratios of 
$\frac{210}{160} \simeq 1.31$
and
$\frac{285}{250} \simeq 1.14$.
We could assert that $\calin$ is fundamentally less 
informative but: 
\begin{enumerate}
\item 
  In 
  \revtexlatexswitch{Sec.~\ref{subsec:fluxwell.othermeasures},
  }{Correlations~with~Other~Health~and~Well-being~Measures in our
    Analysis~and~Results section, }
  we will find that some measures relating to health and well-being correlate
  more strongly with $\calin$ and some with $\calout$;
\item 
  We may adjust the dynamic range of either measure 
  by rescaling, introducing a kind of tunability~\cite{dodds2011e}
  to the instrument (a feature we will reserve for future
  iterations);
  and
\item 
  Because our food phrase database is 
  a factor of 10
  smaller than our activity phrase one,
  revisions of our instrument may elevate
  the power of $\calin$.
\end{enumerate}

\begin{figure}[tp!]
  \begin{center}
    \includegraphics[width=\columnwidth]{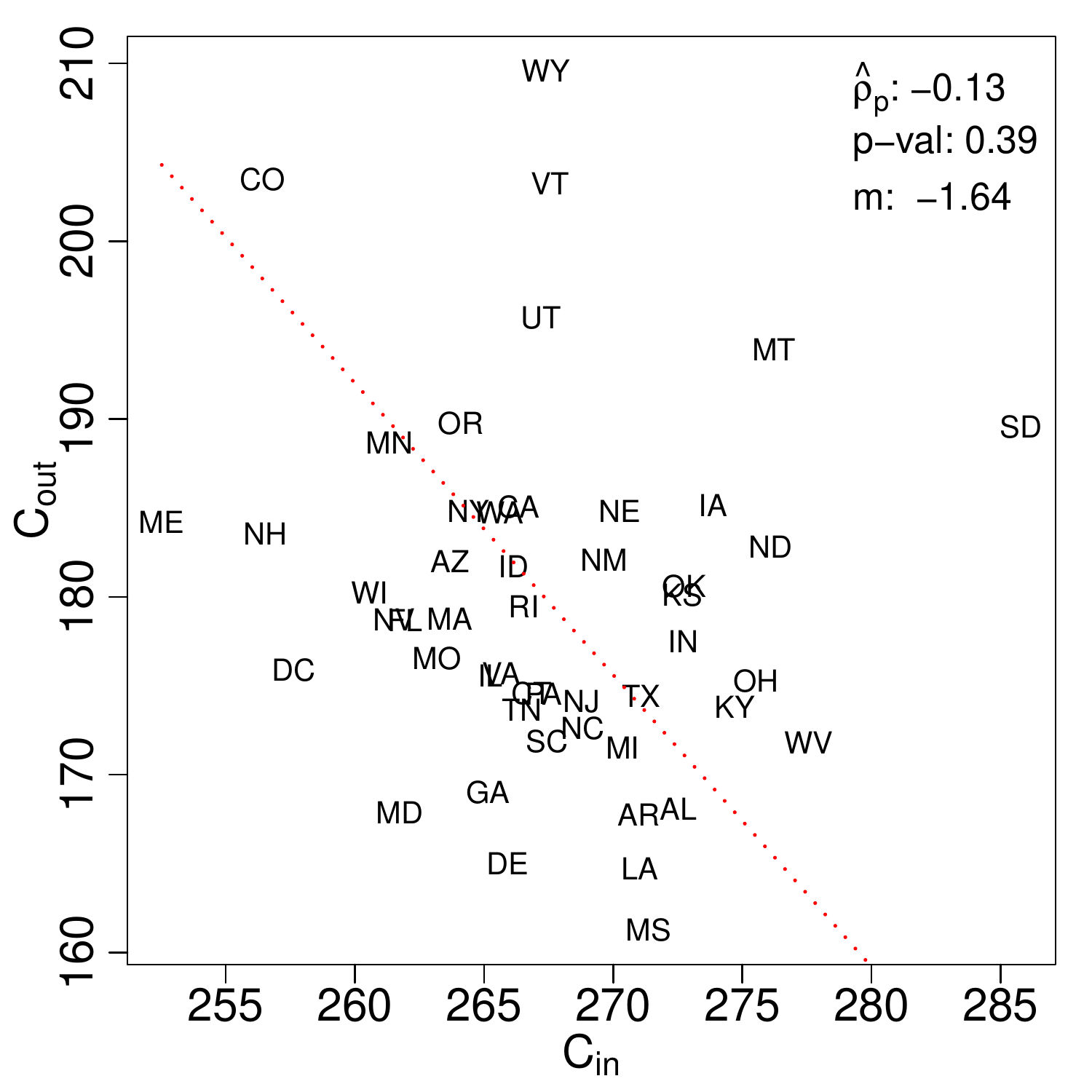}
  \end{center}
  \caption{
    Plots for the contiguous US 
    showing the lack of correlation between
    caloric input $\calin$ and caloric output $\calout$,
    demonstrating their separate value
    as they bear different kinds of information.
    The Pearson correlation coefficient 
    $\rhopearson$
    is -0.13
    and the best line of fit slope is $m=$ -1.64.
            Fig.~\ref{fig:fluxwell.supp.scatterplots} adds
    plots of 
    $\calrat$ as a function of $\calin$ and $\calout$.
  }
  \label{fig:fluxwell.scatterplot_calin_calout}
\end{figure}

To provide some support for point 1, 
we compare $\calout$ and $\calin$ in
Fig.~\ref{fig:fluxwell.scatterplot_calin_calout}
\revtexlatexswitch{(see also Fig.~\ref{fig:fluxwell.supp.scatterplots}).}
{(see also \ref{fig:fluxwell.supp.scatterplots} Fig.).}
Importantly, we see that the two measures are indeed
not well correlated, indicating they contain
different kinds of information
(Pearson correlation coefficient $\rhopearson \simeq 0.13$,
$p$-value = 0.39).
This demonstrates why we might expect 
$\calin$ or $\calout$ to separately correlate
more strongly with other population-level measures,
and justifies forming a dashboard using both $\calin$ and $\calout$
as well the composite measure of $\calrat$.

Regarding point 2 above, we have evidently made
a number of choices in computing $\calin$ and $\calout$
that mean we have already introduced an arbitrary tuning
of the ratio $\calrat$
(e.g., assuming 100 grams of a food and an hour's worth of activity).
Having no principled way of rescaling (i.e., one that is not
a function of the data set being studied), 
we have chosen to leave the measures as computed.
As we discuss later, in future iterations
we envisage for the Caloric Difference version that introducing tunability of the dynamic
ranges of $\calin$ and $\calout$---altering the bias of
the measure toward food or activity---will allow
the Lexicocalorimeter to be refined for 
a range of purposes such as estimating correlates of diabetes levels
versus cancer rates
\revtexlatexswitch{(see Sec.~\ref{subsec:fluxwell.othermeasures}). }{(see Correlations~with~Other~Health~and~Well-being~Measures in Analysis~and~Results). }

\subsection{Rankings for the contiguous US}
\label{subsec:fluxwell.rankings}

\begin{figure*}[tp!]
  \begin{center}
    \includegraphics[width=\textwidth]{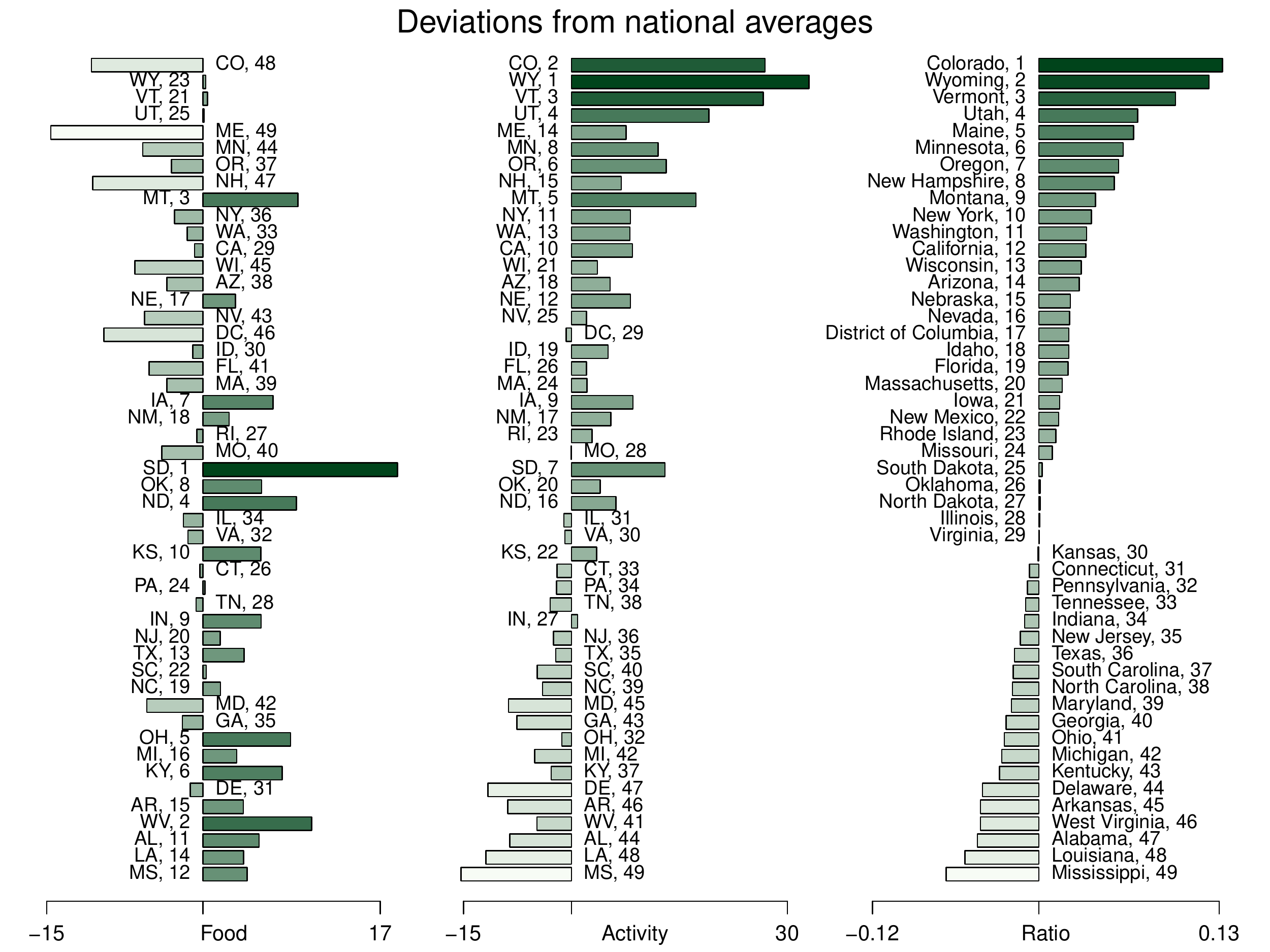}
  \end{center}
  \caption{
    Histograms of 
    caloric intake $\calin$ (food),
    caloric output $\calout$ (activity), 
    and
    caloric ratio $\calrat$
    for the states of the contiguous US,
    all ranked by decreasing $\calrat$.
    Bars indicate the difference in the three quantities from
    the overall average with
    colors corresponding to those used in 
    Figs.~\ref{fig:fluxwell.maps-dominantmodes},
    \ref{fig:fluxwell.maps-diffs},
    and \ref{fig:fluxwell.maps-calrat}.
    We provide the same set of histograms re-sorted by
    $\calin$ and $\calout$
    in
    \revtexlatexswitch{Figs.~\ref{fig:fluxwell.supp.histograms-food}
      and \ref{fig:fluxwell.supp.histograms-activities}.}{
      \ref{fig:fluxwell.supp.histograms-food} Fig.,
      and \ref{fig:fluxwell.supp.histograms-activities} Fig.}
  }
  \label{fig:fluxwell.histograms}
\end{figure*}

Having taken in the maps of our three measures
$\calin$, $\calout$, and $\calrat$,
we now explore the rankings quantitatively, 
first through the histograms 
shown in Fig.~\ref{fig:fluxwell.histograms}.
We order the 48 states and DC by $\calrat$ (rightmost plot)
and all bars are relative to the overall average
of the specific measure.
Numeric rankings for each measure are given next to each bar.
In
\revtexlatexswitch{Figs.~\ref{fig:fluxwell.supp.histograms-food}
  and \ref{fig:fluxwell.supp.histograms-activities},}{
  \ref{fig:fluxwell.supp.histograms-food} Fig.,
  and \ref{fig:fluxwell.supp.histograms-activities} Fig.,}
we present the same histograms re-sorted
respectively by $\calin$ and $\calout$.

As was indicated by our inspection the choropleth maps, 
we do indeed see that $\calrat$ is more strongly 
driven by $\calout$ than $\calin$ due to the former's larger
dynamic range.
The states with the highest values of $\calrat$ achieve
their scores through high levels of $\calout$ but
more variable levels of $\calin$.
Wyoming (23), Vermont (21), and Utah (25)
are all middling in $\calin$ while 
Colorado (48) and Maine (49) have the lowest ranks
for caloric intake.
At the trailing end, we see by contrast
that low activity ranks are coupled with high ranks for caloric intake.

A few of the more anomalous states are both evident
in the $\calin$ and $\calout$ histograms and 
as those appearing farthest away from the best line of fit
in the scatter plot of Fig.~\ref{fig:fluxwell.scatterplot_calin_calout}.
South Dakota has both high values of $\calin$ and $\calout$ 
(ranks of 1 and 7)
that arrange to give it a ranking of 25 for $\calrat$.
Maryland ranking 42nd and 45th in $\calin$ and $\calout$,
is the only state in the `bottom' 10 of both measures.

\subsection{Phrase shifts}
\label{subsec:fluxwell.phraseshifts}

\begin{figure*}[tbp!]
  \begin{center}
    \includegraphics[width=0.95\textwidth]{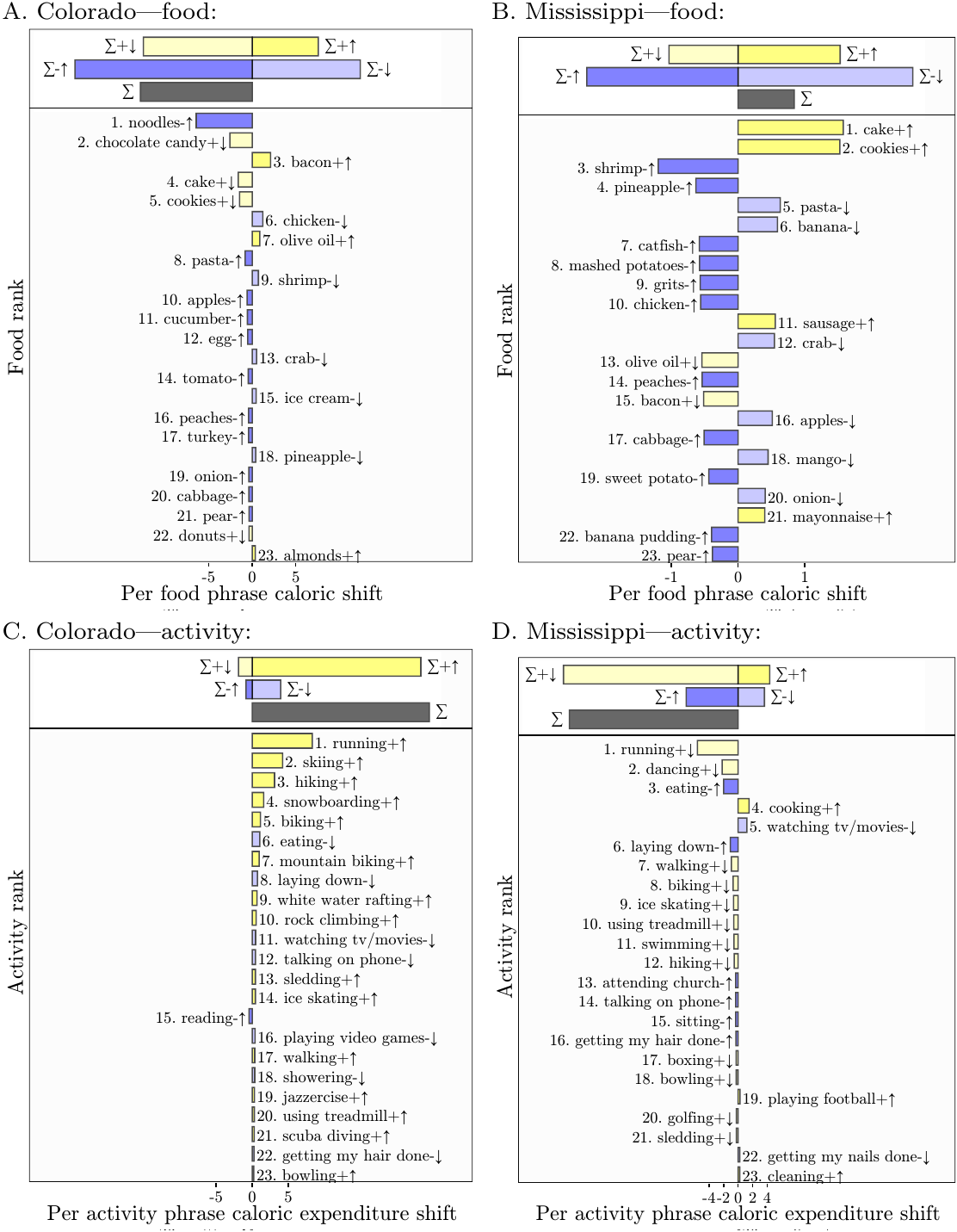} 
          \end{center}
  \caption{
    Phrase shifts showing which food phrases 
    and physical activity phrases
    have the most influence on Colorado and Mississippi's 
    top and bottom ranking for caloric ratio,
    when compared
    with the average for the contiguous United States.
    Note that phrases are lemmas representing phrase categories.
    Overall, Colorado scores lower on Twitter food calories
    (257.4 versus 271.7)      
    and higher 
    on physical activity calories 
    (203.5 versus 161.3)
    than Mississippi.
    We provide interactive phrase shifts 
    as part of the paper's Online
    Appendices at
    \url{http://compstorylab.org/share/papers/alajajian2015a/}
    and at
    \url{http://panometer.org/instruments/lexicocalorimeter}.
    We explain phrase (word) shifts in the main text
    (see Eqs.~\ref{eq:fluxwell.delta} and~\ref{eq:fluxwell.delta-final}),
    and in full depth in~\cite{dodds2011e} 
    and~\cite{dodds2015a}
    and online at
    \url{http://hedonometer.org}~\cite{wordshiftexplanations2014a}.
  }
  \label{fig:fluxwell.wordshiftexample}
\end{figure*}

In our work on measuring happiness, we have developed and extensively used 
``word shifts'' to show which words make a given text appear more positive
than another text in aggregate (see~\cite{dodds2011e} and~\cite{dodds2015a}).
Such visualizations not only provide our necessary test,
but also allow us to draw insight from the lexical tapestry
of texts.
Here, we will explain and use analogously constructed phrase shifts for both 
$\calin$ and $\calout$ to examine the states at the extremes
of our $\calrat$ rankings, 
Colorado and Mississippi.
Interactive food and activity phrase shifts for the 49 regions of the contiguous US
form a central part of our online Lexicocalorimeter:
\url{http://panometer.org/instruments/lexicocalorimeter}.

We start with two texts:
a base ``reference text'' $\textref$,
and a ``comparison text'' $\textcomp$ which we wish to compare
to $\textref$.
In this paper, we will use the Contiguous US as the reference
text (weighting the 
phrase distributions of each state equally),
but in principle any text can be used (e.g., in comparing two
states, one would be selected as a reference).
Our interest is in determining 
which words or phrases most contribute to or go against
the difference in estimated calories.
$\calinoutfn{\textcomp} - \calinoutfn{\textref}$
where i/o stands for in or out.
Following~\cite{dodds2011e} and using~\Req{eq:fluxwell.calin},
we can express the difference as
\begin{eqnarray}
  \lefteqn{\calinoutfn{\textcomp} - \calinoutfn{\textref}} 
  \nonumber \\ 
  & = &
  \sum_{\foodactivityphrase \in \foodactivityphraseset} 
  \calinoutfn{\foodactivityphrase}
  \Big[
    \normfrequency{\foodactivityphrase}{\textcomp}
    -
    \normfrequency{\foodactivityphrase}{\textref}
    \Big]
  \nonumber \\ 
  & = &
  \sum_{\foodactivityphrase \in \foodactivityphraseset} 
  \left[
    \calinoutfn{\foodactivityphrase} - \calinoutfnsup{(\textrm{ref})}
    \right]
  \Big[
    \normfrequency{\foodactivityphrase}{\textcomp}
    -
    \normfrequency{\foodactivityphrase}{\textref}
    \Big].
  \nonumber \\ 
  \label{eq:fluxwell.delta}
\end{eqnarray}
We now have a sum contributions due 
to all phrases. 
We normalize these contributions
as percentages and annotate their structure as follows:
\begin{eqnarray}
  \lefteqn{\delta \calinoutfn{\foodactivityphrase} =}
  \nonumber \\
  & &
  \frac{100}{
    \left|
    \calinoutfnsup{(\textrm{comp})} - \calinoutfnsup{(\textrm{ref})}
    \right| 
  }
  \underbrace{
    \left[
      \calinoutfn{\foodactivityphrase} - \calinoutfnsup{(\textrm{ref})}
      \right]
  }_{+/-}
  \underbrace{
    \left[
      p_{\foodactivityphrase}^{(\textrm{comp})} - p_{\foodactivityphrase}^{(\textrm{ref})}
      \right]
  }_{\uparrow/\downarrow},
  \nonumber \\ 
  \label{eq:fluxwell.delta-final}
\end{eqnarray}
where $\sum_{\foodactivityphrase \in \foodactivityphraseset} 
\delta \calinoutfn{\foodactivityphrase} = \pm 100$.
We use the symbols $+/-$ and $\uparrow/\downarrow$ 
to respectively encode 
whether the calories of a phrase exceed the average
of the reference text,
and whether a phrase is being used more or less
in the comparison text.
We call $\delta \calinoutfn{\foodactivityphrase}$ the 
``per food/activity phrase caloric expenditure shift''.
Finally, we sort phrases by the absolute value of 
$\delta \calinoutfn{\foodactivityphrase}$ to
create each phrase shift.

In Fig.~\ref{fig:fluxwell.wordshiftexample}, we present 
food phrase shifts which help to illustrate why:
\begin{itemize}
\item 
  Colorado ranks 48/49 for caloric input $\calin$
  (Fig.~\ref{fig:fluxwell.wordshiftexample}A),
\item 
  Mississippi ranks 12/49 for caloric input $\calin$
  (Fig.~\ref{fig:fluxwell.wordshiftexample}B),
\item 
  Colorado ranks 2/49 for caloric output $\calout$
  (Fig.~\ref{fig:fluxwell.wordshiftexample}C),
\item 
  and Mississippi ranks 49/49 for caloric output  $\calout$
  (Fig.~\ref{fig:fluxwell.wordshiftexample}D).
\end{itemize}

These shifts display phrases that fall into four categories:
\begin{itemize}[leftmargin=64pt]
\item[+$\uparrow$, yellow:]
  Phrases representing above average quantities (here calories)
  being used more often.
  Examples: 
  ``cookies'' for Mississippi in
  Fig.~\ref{fig:fluxwell.wordshiftexample}B
  and 
  ``rock climbing'' for Colorado in Fig.~\ref{fig:fluxwell.wordshiftexample}C.
\item[-$\downarrow$, pale blue:]
  Phrases representing below average quantities
  being used less often.
  Examples: 
  ``watching tv or movie'' for Mississippi in
  Fig.~\ref{fig:fluxwell.wordshiftexample}B
  and 
  ``laying down'' for Colorado in
  Fig.~\ref{fig:fluxwell.wordshiftexample}C.
\item[+$\downarrow$, pale yellow:]
  Phrases representing above average quantities
  being used less often.
  Examples: 
  ``chocolate candy'' for Colorado in
  Fig.~\ref{fig:fluxwell.wordshiftexample}A
  and 
  ``running'' for Mississippi in Fig.~\ref{fig:fluxwell.wordshiftexample}D.
\item[-$\uparrow$, blue:]
  Phrases representing below average quantities
  being used more often.
  Examples: 
  ``reading'' for Colorado 
  in Fig.~\ref{fig:fluxwell.wordshiftexample}A
  and
  ``catfish'' for Mississippi in
  Fig.~\ref{fig:fluxwell.wordshiftexample}B.
\end{itemize}
Note that depending on the quantity, higher or lower may be ``better''
and the four categories flip signs in their support.
For example, $\calin$ and $\calout$ increase with +$\uparrow$ phrases;
after we examine correlations with health and well-being measures in 
\revtexlatexswitch{Sec.~\ref{subsec:fluxwell.othermeasures}, }{Correlations~with~Other~Health~and~Well-being~measures in Analysis~and~Results, }
we will be able to interpret this as ``bad'' for $\calin$ and ``good'' for
$\calout$.

At the top of each phrase shift, the bars indicate the total
contribution of each of the four types of phrases, and the black
bar the net change.
We see that the four net changes arise in different ways.
\begin{itemize}
\item 
  Fig.~\ref{fig:fluxwell.wordshiftexample}A:
  Colorado is lower than average for $\calin$ largely due
  to tweeting more about relatively low calorie (per 100 grams) foods: ``noodles'',
  ``egg'', ``pasta'', and ``turkey''.  
  We also find
  less tweets about high calorie foods such as ``candy'', ``cake'',
  and ``cookies.''
  Going against these phrases, we see 
  Colorado does tweet relatively more about ``bacon'' and ``olive oil'',
  and less about some relatively lower calorie foods ``chicken'', ``ice
  cream'', ``shrimp'', and ``corn''.
  We note that this does not mean these foods are low calorie
  in absolute terms
  (``ice cream'' is a good example),
  just that 100 grams of them are low calorie in comparison to the US
  baseline.
\item 
  Fig.~\ref{fig:fluxwell.wordshiftexample}B:
  Mississippi almost equally tweets less about a variety
  of low calorie foods, e.g., ``pasta'', ``banana'', and ``crab'' (pale blue bar) 
  while also tweeting more about the complementary range
  of such foods including ``shrimp'', ``peaches'', and ``pineapple'' (dark
  blue bar).
  The modest net gain is mostly due to a small increase
  in tweeting about high calorie foods such 
  as ``cake'', ``cookies'', and ``sausage''.
\item 
  Fig.~\ref{fig:fluxwell.wordshiftexample}C:
  For physical activity, tweets from Colorado show a preponderance
  of relatively high caloric expenditure phrases (+$\uparrow$, yellow)
  including ``running'', ``skiing'', ``hiking'', ``snowboarding'' and
  so on.
  Tweeting less about low effort activities
  is the only other contribution of any substance---Colorado 
  tweets less about ``eating'',
  ``laying down'',
  and ``watching tv or movie''.
\item 
  Fig.~\ref{fig:fluxwell.wordshiftexample}D:
  Mississippi's low ranking in activity is largely
  due to tweeting less about high output activities (+$\downarrow$, pale yellow):
  less ``running'', ``dancing'', ``walking'', and ``biking''.
  The second most important category is an increase
  in low output activity phrases such
  as ``eating'', ``attending church'', and ``talking on the phone.''
\end{itemize}

\revtexlatexswitch{
  In Figs.~\ref{fig:fluxwell.supp.colorado-food},
\ref{fig:fluxwell.supp.mississippi-food},
\ref{fig:fluxwell.supp.colorado-activity},
and
\ref{fig:fluxwell.supp.mississippi-activity},
}{In \ref{fig:fluxwell.supp.colorado-food},
\ref{fig:fluxwell.supp.mississippi-food},
\ref{fig:fluxwell.supp.colorado-activity},
and
\ref{fig:fluxwell.supp.mississippi-activity} Figs.\ }
we complement the four phrase shifts
of Fig.~\ref{fig:fluxwell.wordshiftexample}
by showing the top 23 phrases for 
each of four ways phrases may contribute.
Interactive phrase shifts for all of the contiguous US are housed
at \url{http://panometer.org/instruments/lexicocalorimeter}.

Overall, we find the lexical texture afforded by our phrase shifts 
is generally convincing,
but we expect future improvements in our food and activity data sets
will iron out some oddities (we again use the example of ice cream).
We also note that phrase shifts are very sensitive and
that terms that seem to be being evaluated incorrectly
may easily be removed from the phrase set, and that
doing so will minimally change the overall score for
sufficiently large texts.

\begin{figure*}[tp!]
  \begin{center}
    \includegraphics[width=\textwidth]{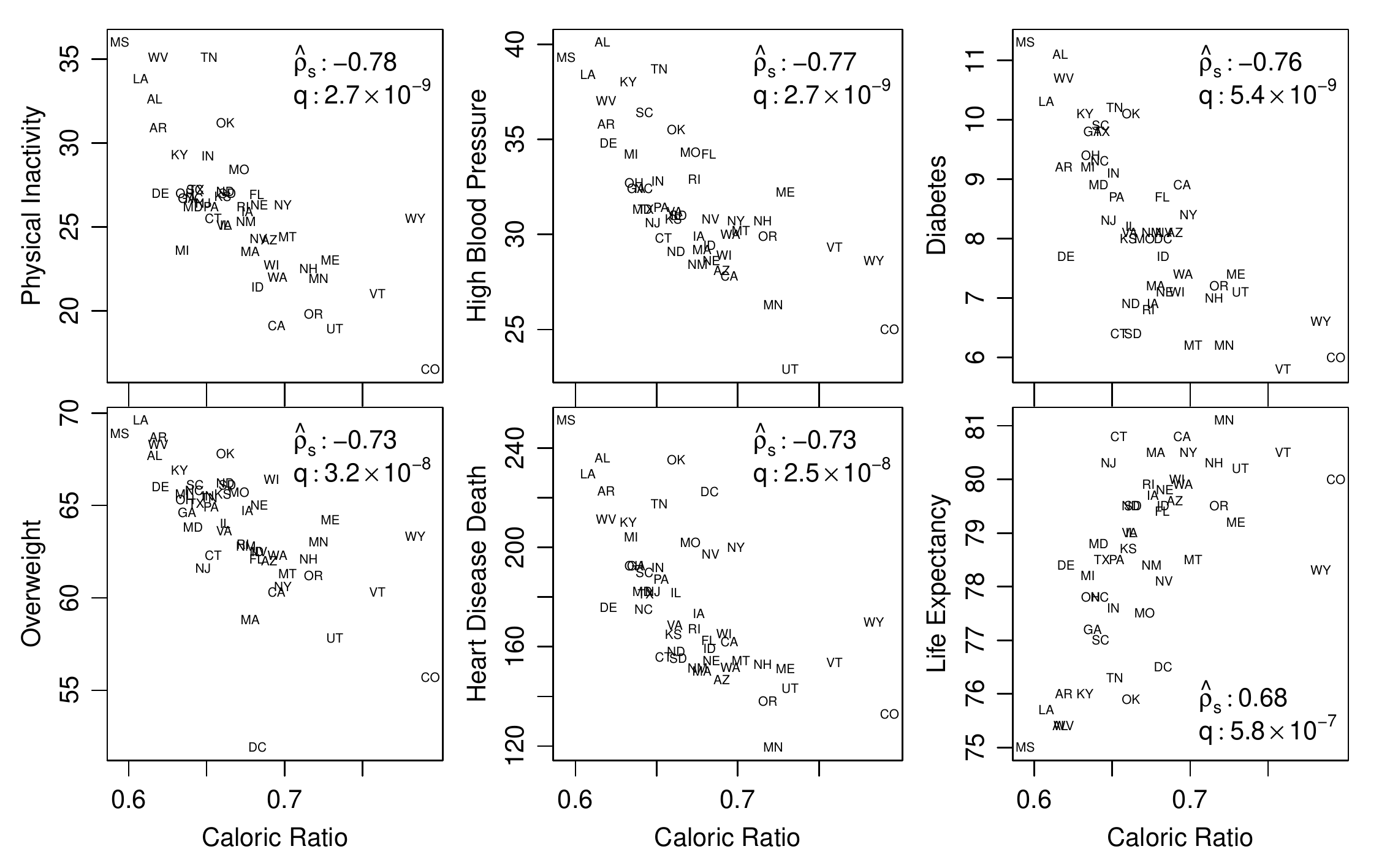}
  \end{center}
  \caption{
    Six demographic quantities compared with
    caloric ratio $\calrat$ for the contiguous US.
    The inset values
    are the Spearman correlation coefficient $\rhospearman$,
    and
    the Benjamini-Hochberg $q$-value.
    See Tab.~\ref{tab:fluxwell.corrTable} for
    a full summary of the 37 demographic quantities
    studied here.
  }
  \label{fig:fluxwell.demog-scatterplots}
\end{figure*}

\subsection{Correlations with other health and well-being measures}
\label{subsec:fluxwell.othermeasures}

\begin{table*}
  \plosoneonly{\begin{adjustwidth}{-4cm}{}}
  \footnotesize
  \begin{tabular}{|l|c|c|c|c|c|c|}
    \hline \multicolumn{1}{|c|}{\textbf{Health and/or well-being
        quantity}}
    & \begin{tabular}{@{}c@{}} $\rhospearman$ for \\ $\calrat$ 
      \end{tabular} 
    & \textbf{$q$-val}
    & \begin{tabular}{@{}c@{}}$\rhospearman$ for \\ $\calin$ 
      \end{tabular} & \textbf{$q$-val} & 
    \begin{tabular}{@{}c@{}}$\rhospearman$ for \\ 
      $\calout$ \end{tabular} & 
    \textbf{$q$-val} \\\hline
    \textbf{1.} \% no physical activity in past 30 days \cite{america} & -0.78 & $2.73\times 10^{-09}$ & 0.58 & $5.67\times 10^{-05}$ & -0.66 & $1.51\times 10^{-06}$ \\\hline
    \textbf{2.} \% have been physically active in past 30 days \cite{america} & 0.78 & $2.73\times 10^{-09}$ & -0.57 & $6.53\times 10^{-05}$ & 0.67 & $1.24\times 10^{-06}$ \\\hline
    \textbf{3.} \% high blood pressure \cite{america} & -0.77 & $2.73\times 10^{-09}$ & 0.32 & $4.05\times 10^{-02}$ & -0.78 & $2.73\times 10^{-09}$ \\\hline
    \textbf{4.} Adult diabetes rate \cite{cdc} & -0.76 & $5.44\times 10^{-09}$ & 0.29 & $6.09\times 10^{-02}$ & -0.77 & $2.73\times 10^{-09}$ \\\hline
    \textbf{5.} CNBC quality of life ranking \cite{cnbc} & -0.76 & $6.75\times 10^{-09}$ & 0.28 & $7.34\times 10^{-02}$ & -0.77 & $3.60\times 10^{-09}$ \\\hline
    \textbf{6.} \% adult overweight/obesity \cite{kff} & -0.73 & $3.16\times 10^{-08}$ & 0.55 & $1.41\times 10^{-04}$ & -0.59 & $3.07\times 10^{-05}$ \\\hline
    \textbf{7.} Heart disease death rate \cite{kff} & -0.73 & $2.50\times 10^{-08}$ & 0.34 & $2.80\times 10^{-02}$ & -0.73 & $2.30\times 10^{-08}$ \\\hline
    \textbf{8.} \% adult obesity \cite{cdc} & -0.72 & $4.30\times 10^{-08}$ & 0.53 & $2.26\times 10^{-04}$ & -0.59 & $2.96\times 10^{-05}$ \\\hline
    \textbf{9.} Gallup Wellbeing score \cite{gallup} & 0.72 & $4.69\times 10^{-08}$ & -0.31 & $4.43\times 10^{-02}$ & 0.73 & $3.99\times 10^{-08}$ \\\hline
    \textbf{10.} America's Health Rankings, overall \cite{america} & -0.72 & $4.10\times 10^{-07}$ & 0.43 & $4.74\times 10^{-03}$ & -0.67 & $2.77\times 10^{-06}$ \\\hline
    \textbf{11.} Life expectancy at birth \cite{kff}  & 0.68 & $5.81\times 10^{-07}$ & -0.4 & $6.91\times 10^{-03}$ & 0.65 & $2.64\times 10^{-06}$ \\\hline
    \textbf{12.} \% who eat fruit less than once a day \cite{produce} & -0.67 & $1.20\times 10^{-06}$ & 0.61 & $1.39\times 10^{-05}$ & -0.51 & $5.35\times 10^{-04}$ \\\hline
    \textbf{13.} \% child overweight/obesity \cite{kff} & -0.64 & $3.53\times 10^{-06}$ & 0.27 & $7.55\times 10^{-02}$ & -0.64 & $3.20\times 10^{-06}$ \\\hline
    \textbf{14.} \% who eat vegetables less than once a day \cite{produce} & -0.61 & $1.39\times 10^{-05}$ & 0.51 & $5.33\times 10^{-04}$ & -0.46 & $1.57\times 10^{-03}$ \\\hline
    \textbf{15.} Median daily intake of fruits \cite{produce} & 0.6 & $1.98\times 10^{-05}$ & -0.62 & $8.33\times 10^{-06}$ & 0.41 & $5.37\times 10^{-03}$ \\\hline
    \textbf{16.} Smoking rate \cite{kff}  & -0.59 & $2.96\times 10^{-05}$ & 0.51 & $5.26\times 10^{-04}$ & -0.48 & $1.08\times 10^{-03}$ \\\hline
    \textbf{17.} Median household income \cite{kff} & 0.51 & $5.55\times 10^{-04}$ & -0.53 & $3.27\times 10^{-04}$ & 0.4 & $8.38\times 10^{-03}$ \\\hline
    \textbf{18.} Median daily intake of vegetables \cite{produce} & 0.5 & $6.10\times 10^{-04}$ & -0.56 & $7.44\times 10^{-05}$ & 0.31 & $4.36\times 10^{-02}$ \\\hline
    \textbf{19.} \% high cholesterol \cite{america} & -0.49 & $8.11\times 10^{-04}$ & 0.23 & $1.45\times 10^{-01}$ & -0.48 & $9.05\times 10^{-04}$ \\\hline
    \textbf{20.} Brain health ranking \cite{brain} (lower is better) & -0.49 & $8.11\times 10^{-04}$ & 0.62 & $1.39\times 10^{-05}$ & -0.29 & $5.70\times 10^{-02}$ \\\hline
    \textbf{21.} \% with bachelor's degree or higher \cite{census} & 0.46 & $1.57\times 10^{-03}$ & -0.54 & $1.66\times 10^{-04}$ & 0.33 & $2.82\times 10^{-02}$ \\\hline
    \textbf{22.} Colorectal cancer rate \cite{cdc} & -0.44 & $4.09\times 10^{-03}$ & 0.53 & $3.59\times 10^{-04}$ & -0.27 & $8.25\times 10^{-02}$ \\\hline
    \textbf{23.} US Census Gini index score \cite{gini} (lower is better) & -0.42 & $5.37\times 10^{-03}$ & -0.03 & $8.42\times 10^{-01}$ & -0.5 & $5.55\times 10^{-04}$ \\\hline
    \textbf{24.} Avg \# poor mental health days, past 30 days \cite{america} & -0.42 & $5.37\times 10^{-03}$ & 0.12 & $4.80\times 10^{-01}$ & -0.48 & $1.06\times 10^{-03}$
    \\\hline \hline
    \textbf{25.} Neuroticism Big Five personality trait \cite{rentfrow} & -0.38 & $1.09\times 10^{-02}$ & 0.2 & $2.03\times 10^{-01}$ & -0.37 & $1.44\times 10^{-02}$ \\\hline
    \textbf{26.} Binge drinking rate \cite{america} & 0.37 & $1.46\times 10^{-02}$ & -0.15 & $3.56\times 10^{-01}$ & 0.41 & $5.84\times 10^{-03}$ \\\hline
    \textbf{27.} Avg \# poor physical health days, past 30 days \cite{america} & -0.35 & $2.34\times 10^{-02}$ & 0.19 & $2.19\times 10^{-01}$ & -0.38 & $1.13\times 10^{-02}$ \\\hline
    \textbf{28.} Farmers markets per 100,000 in pop. \cite{produce} & 0.34 & $2.72\times 10^{-02}$ & 0.06 & $7.17\times 10^{-01}$ & 0.42 & $5.14\times 10^{-03}$
    \\\hline \hline
    \textbf{29.} Strolling of the Heifers locavore score (lower is better) \cite{stroll} & -0.29 & $5.86\times 10^{-02}$ & -0.3 & $5.41\times 10^{-02}$ & -0.45 & $2.94\times 10^{-03}$ \\\hline
    \textbf{30.} Extraversion Big Five personality trait \cite{rentfrow} & -0.28 & $6.94\times 10^{-02}$ & 0.03 & $8.42\times 10^{-01}$ & -0.29 & $5.63\times 10^{-02}$ \\\hline
    \textbf{31.} \% schools offering fruit/veg at celebrations \cite{produce} & 0.24 & $1.31\times 10^{-01}$ & -0.46 & $1.96\times 10^{-03}$ & 0.05 & $7.90\times 10^{-01}$ \\\hline
    \textbf{32.} Openness Big Five personality trait \cite{rentfrow} & 0.23 & $1.31\times 10^{-01}$ & -0.5 & $6.11\times 10^{-04}$ & 0.04 & $8.10\times 10^{-01}$ \\\hline
    \textbf{33.} \% cropland harvested for fruits/veg \cite{produce} & 0.19 & $2.34\times 10^{-01}$ & -0.62 & $1.37\times 10^{-05}$ & -0.04 & $8.10\times 10^{-01}$ \\\hline
    \textbf{34.} Conscientiousness Big Five personality trait \cite{rentfrow} & -0.12 & $4.81\times 10^{-01}$ & 0.2 & $2.10\times 10^{-01}$ & -0.05 & $7.93\times 10^{-01}$ \\\hline
    \textbf{35.} \% census tracts, healthy food retailer within 1/2 mile \cite{produce} & -0.03 & $8.44\times 10^{-01}$ & -0.52 & $3.68\times 10^{-04}$ & -0.24 & $1.31\times 10^{-01}$ \\\hline
    \textbf{36.} George Mason overall freedom ranking \cite{freedom} (lower is freer) & -0.03 & $8.42\times 10^{-01}$ & -0.11 & $5.15\times 10^{-01}$ & -0.1 & $5.64\times 10^{-01}$ \\\hline
    \textbf{37.} Agreeableness Big Five personality trait \cite{rentfrow} & -0.01 & $9.61\times 10^{-01}$ & 0.22 & $1.50\times 10^{-01}$ & 0.08 & $6.47\times 10^{-01}$ 
    \\\hline
  \end{tabular}
  \caption{
        Spearman correlation coefficients, $\rhospearman$, and Benjamini-Hochberg $q$-values for
    caloric input $\calin$,
    caloric output $\calout$, 
    and caloric ratio $\calrat = \calout/\calin$
    and
    demographic, 
    data related to food and physical activity, Big Five
    personality traits~\cite{rentfrow}, health and well-being rankings by state, and
    socioeconomic status, correlated, ordered from strongest to weakest Spearman
    correlations with caloric ratio.
    The two breaks in the table indicate significance levels of 0.01 and
    0.05
    for the Benjamini-Hochberg $q$ of $\calrat$,
    corresponding to the first 24 health and/or well-being quantities and
    then the next four, numbers 25 to 28.
    The bottom 9 quantities were not significantly correlated
    with $\calrat$ according to our tests.
    \revtexlatexswitch{Tabs.~\ref{tab:fluxwell.supp.corrTableLIQ},
      \ref{tab:fluxwell.supp.corrTableNOLIQ_DIF},
      and
      \ref{tab:fluxwell.supp.corrTableLIQ_DIF}}{
      \ref{tab:fluxwell.supp.corrTableLIQ},
      \ref{tab:fluxwell.supp.corrTableNOLIQ_DIF},
      and
      \ref{tab:fluxwell.supp.corrTableLIQ_DIF}
      Tabs.\ }
                      present the same analysis for caloric measures including phrases
                      representing liquids,
                      and 
                      for the difference $\caldiff(\alpha) = \alpha\calout - (1-\alpha)\calin$,
                      both without and with liquids included.
  }
  \label{tab:fluxwell.corrTable}
  \plosoneonly{\end{adjustwidth}}
\end{table*}

We now turn to a suite of statistical comparisons
between our three measures---caloric input, caloric output,
and caloric ratio---and a collection of 
demographic, behavioral, health, and psychological quantities.

We use Spearman's correlation coefficient $\rhospearman$ to examine relationships between 
$\calin$, $\calout$, and $\calrat$ and 37 variables
variously relating to food and physical activity, 
``Big Five'' personality traits,
and health and well-being rankings (a total of 111 comparisons)
\cite{america,brain,cdc,census,cnbc,freedom,gallup,gini,kff,produce,rentfrow,stroll}.
To correct for multiple comparisons, we calculate the
$q$-value for each correlation coefficient using the
Benjamini-Hochberg step-up procedure~\cite{benjamini1995}
(the $q$-value is to be interpreted in the same way as a $p$-value).
We then consider correlations in reference
to the standard significance levels of 0.01 and 0.05.

We must first acknowledge that many of the variables we test 
against our measures are highly correlated with each other.
The food and physical activity-related variables are in the areas of
physical activity levels, produce intake and availability rates
(including trends in public schools), chronic disease rates, and rates
of unhealthy habits.  
Many of these variables are well known to be
influenced by diet and physical activity (e.g., obesity
rates~\cite{cdc}), 
and others may be less directly related (e.g., percent of
cropland in each state harvested for fruits and
vegetables~\cite{produce}).  

To give some grounding for the full set of comparisons,
we show in Fig.~\ref{fig:fluxwell.demog-scatterplots}
how six demographic quantities vary with caloric ratio $\calrat$.
We see strong correlations with $|\rhospearman| \ge 0.68$,
and the highest value for Benjamini-Hochberg $q$-value is 5.8$\times$$10^{-7}$.

We present a summary of all results in Tab.~\ref{tab:fluxwell.corrTable}
where we have ordered and numbered demographic quantities in terms of
ascending  
Benjamini-Hochberg $q$-values for $\calrat$.
For comparison and to further demonstrate 
the robustness of our approach, in
\revtexlatexswitch{Tabs.~\ref{tab:fluxwell.supp.corrTableLIQ},
  \ref{tab:fluxwell.supp.corrTableNOLIQ_DIF},
  and
  \ref{tab:fluxwell.supp.corrTableLIQ_DIF}),}{
  (see 
  \ref{tab:fluxwell.supp.corrTableLIQ},
  \ref{tab:fluxwell.supp.corrTableNOLIQ_DIF},
  and
  \ref{tab:fluxwell.supp.corrTableLIQ_DIF}
  Tables,
}
we reproduce the same analysis 
with the inclusion of liquids and for
a differential measure 
$\caldiff(\alpha) = \alpha\calout - (1-\alpha)\calin$,
both with and without liquids.
Here, we choose to set the effective means 
of $\calout$ and $\calin$ equal 
across the statewide averages
(i.e., 
$
\alpha 
\tavg{\calout}
=
(1-\alpha)
\tavg{\calin}$),
resulting in $\alpha = 0.598$.
Overall, we find little variation in our results 
whether we use $\calrat$ and $\caldiff(0.598)$.

Surveying the health-based demographics,
we found $\calrat$  
was significantly correlated with 
all chronic
disease-related rates we tested against (high blood pressure (\#3),
adult diabetes (\#4), adult overweight and
obesity (\#6), heart disease deaths (\#7),
adult obesity (\#8), childhood overweight and obesity
(\#13), high cholesterol (\#19), and colorectal cancer (\#22)).
All of these but colorectal
cancer rate were also significantly correlated with $\calout$.

Caloric input $\calin$ results were more mixed.  
Chronic disease-related rates were also significantly correlated with 
$\calin$, with the exception of adult diabetes, childhood overweight and obesity,
and high cholesterol, after correcting for multiple comparisons.  

The variables relating to
unhealthy habits (smoking (\#16) and binge drinking rates (\#26)) both
correlated significantly with all three of our measures with
the one exception of binge drinking and caloric input.
The direction of correlations for these two habits 
are opposite each other (e.g., negative for smoking and $\calrat$, 
positive for binge drinking and $\calrat$), consistent with 
recent work on alcohol consumption~\cite{french2009a}.

The two variables relating to physical activity rates (percent of
population that has had no physical activity in past 30 days (\#1),
and percent of population that has been physically active in past 30 days (\#2))
correlated significantly with all three of our measures.
The two measures relating to rates of physical and mental health
(average number of poor mental health days in past 30 days (\#24), and
average number of poor physical health days in past 30 days (\#27))
correlated significantly with both $\calout$ and $\calrat$,
but did not correlate significantly with $\calin$.

The four variables relating to fruit and vegetable consumption rates
all correlated significantly with all three of our measures.
The variables relating to presence of produce in the state (percent of
cropland in each state harvested for fruits and vegetables (\#33),
percent of census tracts with a healthy food retailer within one-half
mile (\#35), and percent of schools offering fruits and vegetables at
celebrations (\#31)) were significantly correlated with $\calin$
but were not correlated with $\calout$ or $\calrat$.
Variables relating to local food (number of farmers markets per
100,000 people (\#28) and Strolling of the Heifers locavore
score (\#29)) were not significantly correlated with $\calin$,
but were significantly correlated with $\calout$.

Our health and well-being ranking variables included 
the CNBC quality of life ranking (\#5), 
Gallup Wellbeing ranking (\#9), 
America's Health Ranking overall state rank (\#10), 
life expectancy ranking (\#11), 
Brain Health ranking (\#20), 
Gini index score (\#23), and
George Mason's overall freedom ranking (\#36).  
Caloric ratio
correlated with all of these variables except for George Mason's
freedom ranking (which did not correlate with any of our three
measures).  $\calout$ correlated significantly with all of these
measures except for the Brain Health ranking and the freedom ranking.
caloric input $\calin$ did not correlate significantly with the CNBC quality
of life ranking, Gini index score, or freedom ranking.

Regarding correlations with the \textit{Big Five} personality traits,
Pesta \etal\ noted that ``Neuroticism...emerged as the only
consistent Big Five predictor of epidemiologic outcomes (e.g., rates
of heart disease or high blood pressure) and health-related behaviors
(e.g., rates of smoking or exercise)''~\cite{pesta}.  
Additionally,
``neuroticism correlates with many health-related variables, including
depression and anxiety disorders, mortality, coping skill, death from
cardiovascular disease, and whether one smokes tobacco''~\cite{pesta}.
Here, in keeping with these observations, we found that neuroticism
(\#25) was 
indeed the only \textit{Big Five}
personality trait that correlated significantly and negatively  
with caloric ratio.

We also tested our three measures against two measures of
socioeconomic status---median income (\#17) and percent of state with a
bachelor's degree or higher level of education (\#21)---and found
these correlations were significant for all three of our measures.

\section{Concluding Remarks}
\label{sec:fluxwell.conclusion}

\begin{figure*}[tp!]
  \begin{center}
    \includegraphics[width=\textwidth]{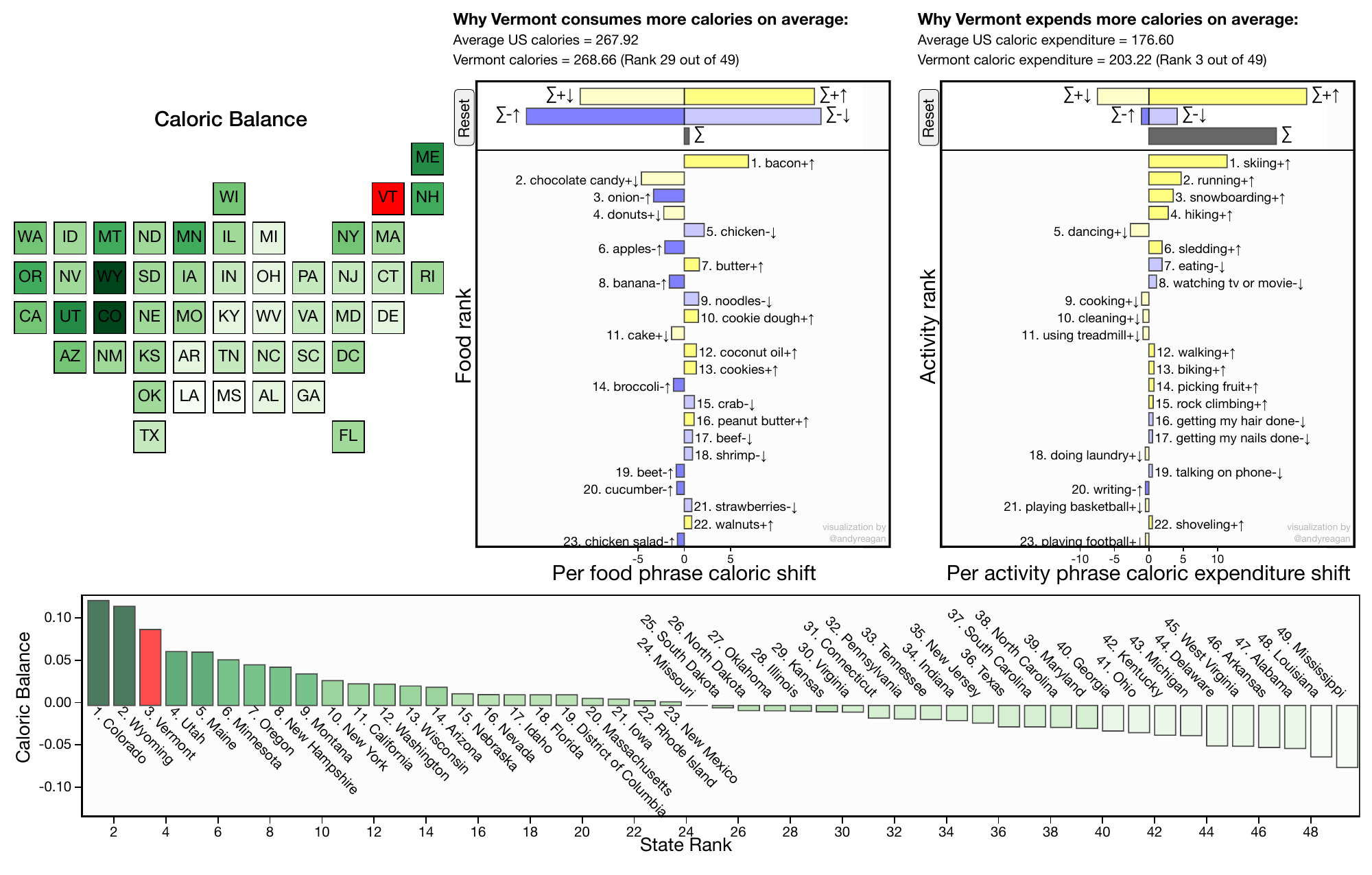}
  \end{center}
  \caption{
    Screenshot of the interactive dashboard for our prototype
    Lexicocalorimeter site (taken 2015/07/03).
    An archived development version can be found
    as part of our paper's Online Appendices at
    \url{http://compstorylab.org/share/papers/alajajian2015a/maps.html},
    and a full dynamic implementation will be part of our
    Panometer project at
    \url{http://panometer.org/instruments/lexicocalorimeter}.
    See
    \url{https://github.com/andyreagan/lexicocalorimeter-appendix}
    for source code.
  }
  \label{fig:fluxwell.lexicocalorimeter}
\end{figure*}

Our Lexicocalorimeter
has thus, when applied to Twitter,
proved to find and demonstrate 
a range of strong, commonsensical 
patterns and correlations for the contiguous US.
We invite the reader to explore our online instrument,
a screenshot of which is shown in Fig.~\ref{fig:fluxwell.lexicocalorimeter}.

Given the complex relationships between health,
well-being, happiness, and various measures of socioeconomic status,
it is rather difficult to say that we are only measuring health or
only measuring well-being.  We are also measuring socioeconomic
status to some extent.  However, the correlations between caloric
ratio and measures of socioeconomic status are not as strong as the
correlation of caloric ratio with many of the other measures.
Given the above, we believe that the caloric content of
tweets can be used successfully, along with other well-being and
quality of life measures, to help gauge overall well-being in a population.

There are many potential forward directions. 
A promising avenue is to 
incorporate tunability to the Lexicocalorimeter by manipulating its dynamic range.
While we chose the caloric ratio $\calrat$ for its generality in the main body of this work, 
there is more flexibility in the measurement of caloric difference:
$\caldiff(\alpha) = \alpha \calout - (1-\alpha) \calin$.
Though a universal approach is unclear 
($\alpha$ should be independent of the particular data set being studied), 
we may profit from the versatility of $\caldiff(\alpha)$ when focusing on a single demographic. 
For example, if we are interested in diabetes rates, 
we could tune the instrument to 
obtain the best correlation with known levels, 
and thereby create a real-time estimator. 
To do so, we would tune $\alpha$ and 
find the value that gives the highest correlation 
between $\caldiff(\alpha)$ and diabetes rates 
for a given set of populations. 
Of course, we could use a ``black box'' method to
generate a more optimal fit, 
but in basing our instrument on food and activity words,
we have a far more principled approach 
that grants us the opportunity not just to mimic 
but to understand and explain patterns that we find. 
In particular, our word shifts will be of great use
in showing why our hypothetical estimate of diabetes 
is varying across populations.

We fully recognize
that the Twitter population is not the same as the general population;
Twitter users differ from the general population in terms of race,
age, and urbanity~\cite{pew}.  
However, we currently have no reliable
way to know, for example, the true age, race, gender, and education
level of individual users and as such, are not able to adjust for
these factors.  
While we were able to vet our food and physical
activity lists to some extent (as described in Methods and Materials), we could
not realistically go through every tweet to be certain that the phrase
was being used in the way that we thought.  
We realize that even if
the phrases are being used as we imagine, it does not necessarily mean
that the person who tweeted actually performed the physical activity
or ate the tweeted-about food (West \etal\ address a similar
issue in inferring food consumption from accessing recipes online~\cite{west2013}).

We also currently do not know at what point our metric breaks down at smaller
time scales (e.g., months or weeks) or for smaller spatial regions
(e.g., city or county) level.  Our preliminary research shows that the
physical activity metric on its own may be quite effective at the city
level, but the food measure may not be accurate on a smaller scale.
We have also found the physical activity list to be robust to random
partitioning~\cite{williams2015a}, whereas the food list was not.  
We believe that these preliminary findings may be due to several factors:
(a) the size of the food list (just over 1400 phrases) is much smaller
than the physical activity phrase list (just over 13,400 phrases); 
(b)
there are generally more tweets about physical activities in our
list than the foods in our food list; 
and (c) the amount of data
within a city may not be a large enough sample for any food-based
Twitter metric.
We note that we have not tried using the
metric on counties or Census block or tract groups, and it may be that
these are more conducive to the metric.

We propose to use crowdsourcing as a way to build a more comprehensive
food phrase list that includes commonly eaten foods with brand names
as well as food slang that we did not capture here.
Ideally, we would arrive at a food phrase database similar in
scale to that of our existing physical activity phrase list.
However we move forward, we believe it is clear that the 
Lexicocalorimeter we have designed and implemented is already of some potency
and may be improved substantively in the future.

\section{Methods and Materials}
\label{sec:fluxwell.methods}

In order to attempt to estimate the ``caloric content''
of text-extracted phrases~\cite{williams2015a} relating to food
(caloric input) and physical activity (caloric output), we needed
comprehensive lists of foods and physical activities and their
respective caloric content and expenditure information.
Here, we explain in detail how we constructed these phrase lists 
and assigned calories to each phrase.

In \revtexlatexswitch{dataset S1}{S1 dataset} (\url{https://dx.doi.org/10.6084/m9.figshare.4530965.v1}), 
we provide message IDs for all tweets that are part of our study,
and we make both this dataset and other material and visualizations
available at the paper's Online Appendices
(\url{http://compstorylab.org/share/papers/alajajian2015a/},
and as part of our Panometer project
at \url{http://panometer.org/instruments/lexicocalorimeter}.
We have drawn on Twitter's Gardenhose API which has been provided
to the Computational Story Lab by Twitter.

\subsection{Calorie estimates for phrases}
\label{subsec:fluxwell.methods-phrases-to-calories}

We used the USDA National Nutrient Database~\cite{usda} to approximate
the caloric content of foods, and the Compendium of Physical Activities
from Arizona State University and the National Cancer
Institute~\cite{compendium} to approximate average Metabolic
Equivalent of Tasks (METs) for physical activities, which we converted
to calories expended per hour of activity~\cite{compendium}.
Because the foods listed in the USDA National Nutrient Database are
not described in a way that people talk about food, we created a list
of food phrases used on Twitter by starting with a kernel of basic
food terms from the USDA's MyPlate website's food group
pages~\cite{myplate}.
If the food phrase was not specific, such as ``cereal'', we chose the
most popular version of that food in the United States via an informal
Google search at the time of the study (in this instance, Cheerios).
If a brand name food was not in the USDA National Nutrient Database,
we chose the closest match we could find.  
(Please note that this means that data in appendix may be inaccurate
when searching brand name items.)

This approach yielded examples of foods in the food groups of fruits,
vegetables, grains, proteins, dairy, oils, solid fats, and 
``empty calories'' (e.g., junk food), 
and built up a list of nearly 1400 food phrases used on Twitter.
For the main results we present in this study, 
we did not include drinks or soups (liquids) in our list.
We found there is very little change
in our findings when liquids are included,
as we discuss below, 
and we have omitted them at present both for 
simplicity and because
we were not satisfied with a straightforward
way of balancing liquid and solid nutrition estimates.
Note that we have included ice creams, oils, and some other items 
that may act as liquids,
and these could be separated out for future versions of our instrument.

For physical activity, we used the physical activities listed in the
Compendium to build up a list of nearly 14,000 physical activity
phrases used on Twitter.
The order of magnitude of difference between the length of the two
lists exists because of the difference in the number of terms that
went into creating each list and the rates at which people tweet about
foods vs. physical activities.

\subsection{Phrase extraction}
\label{subsec:fluxwell.methods-phrase-extraction}

A major obstacle to the development of the food and physical activity
lists is the determination of those phrases used by individuals that
most accurately represent a food or physical activity.  Various
methods exist which may help one ascertain information about the
frequency of usage of higher-order lexical units~\cite{williams2015a}.
However, we require one that not only determines reasonable estimates
of frequency of usage, but further, does so with nuance regarding
context.  For example, one should not count the phrase ``apple'' as
having occurred if it appeared within a larger phrase that was
recognized as meaningful, such as ``you're the apple of my eye.''
To accomplish these goals, we define a low-assumption
text segmentation algorithm, which we refer to as \emph{serial
  partitioning}.

Serial text partitioning is a greedy algorithm
\revtexonly{(see Alg.~\ref{fluxwell:serialPartitioning})}
for finding distinct, coherent subsequences (phrases) within a sequence (clause).
It relies on the directionality of a sequence,
and so is particularly adept for processing text
into multi-word expressions for many modern languages.
The algorithm relies on an objective function,
which we will generally refer to as $\mathcal{L}$.
At a high level, the algorithm seeks to find
find the largest subsequences possible,
following a chain of optimizing, growing subsequences.

In the context of this article, we define $\mathcal{L}$ relative to a
text $T$ as follows, providing pseudocode below.
First, let $f:S\rightarrow \mathbb{R}^{\geq 0}$ be the random partition frequency
function \cite{williams2015a} under the pure random partition probability ($q = \frac{1}{2}$)
for the text $T$. We then apply the model of context developed in \cite{williams2015c}
under the parameterization $q = 1$, so that a given phrase $s$ is a member of $\ell(s)$ contexts
$\mathcal{C}_s$
(e.g., the phrase $s = (New, York, City)$ is a member of three contexts,
labeled $\mathcal{C}_s = \{(*, York, City)$, $(New, *, City)$, and $(New, York, *)\}$).
Then for $C\in\mathcal{C}_s$,
we consider the context-local likelihood probabilities:
\begin{equation}
  P(s\mid C) = \frac{f(s)}{\underset{t\in C}{\sum}f(t)},
  \label{eq.fluxwell:PsC}
\end{equation}
and prescribe to $s$ the likelihood-minimizing context
\begin{equation}
  C_s = \underset{C\in\mathcal{C}_s}{\text{argmin}}(P(s\mid C)),
  \label{eq.fluxwell:Cs}
\end{equation}
which chooses the context-pattern that is most prevalent in $T$.
The objective function for this instantiation of serial partitioning
is then defined as
\begin{equation}
  \mathcal{L}(s) = P(s\mid\,C_s),
  \label{eq.fluxwell:Ls}
\end{equation}
and referred to as the \emph{local likelihood} of a phrase $s$.

\revtexlatexswitch{
\begin{algorithm}[H]
  \begin{algorithmic}[1]
    \Procedure{SerialTextPartitioning}{t}
    \State $\mathcal{P} \gets (\cdot)$ \Comment{init. the partition.}
    \State $s \gets (\cdot)$ \Comment{init. the phrase.}
    \For {$i\in(1, \cdots, \ell(t))$}
    \If {$\mathcal{L}(s^\frown t_i) > \mathcal{L}(s)$}
    \State $s \gets s^\frown t_i$
    \Else
    \State $\mathcal{P}\gets \mathcal{P}^\frown s$
    \State $s \gets t_i$
    \EndIf
    \EndFor
    \State \textbf{return} $\mathcal{P}$
    \EndProcedure
  \end{algorithmic}
  \caption{
    Serial text partitioning of a (left-to-right) directional clause,
    given an objective function $\mathcal{L}:S\rightarrow\mathbb{R}^{\geq0}$
    (whose maximization is desired, in this case)
    that is zero on the empty phrase $(\cdot)$,
    and a clause $t = (t_1,\cdots, t_{\ell(t)})$,
    consisting of $\ell(t)$ words.
    Note that for any $a,b\in S$,
    $a^\frown b = (a_1,\cdots,a_{\ell(a)},b_1,\cdots b_{\ell(b)})$
    denotes the concatenation of phrases, and that for convenience,
    a single sequence element, $a_i$, may be treated as sequence of one term, $(a_i)$.
  }
  \label{fluxwell:serialPartitioning}
\end{algorithm}
}{

  An outline of serial text partitioning of a (left-to-right) directional clause,
  given an objective function $\mathcal{L}:S\rightarrow\mathbb{R}^{\geq0}$
  (whose maximization is desired, in this case)
  that is zero on the empty phrase $(\cdot)$,
  and a clause $t = (t_1,\cdots, t_{\ell(t)})$,
  consisting of $\ell(t)$ words is as follows:

  \begin{algorithmic}[1]
    \Procedure{SerialTextPartitioning}{t}
    \State $\mathcal{P} \gets (\cdot)$ \Comment{init. the partition.}
    \State $s \gets (\cdot)$ \Comment{init. the phrase.}
    \For {$i\in(1, \cdots, \ell(t))$}
    \If {$\mathcal{L}(s^\frown t_i) > \mathcal{L}(s)$}
    \State $s \gets s^\frown t_i$
    \Else
    \State $\mathcal{P}\gets \mathcal{P}^\frown s$
    \State $s \gets t_i$
    \EndIf
    \EndFor
    \State \textbf{return} $\mathcal{P}$
    \EndProcedure
  \end{algorithmic}
  
  Note that for any $a,b\in S$,
  $a^\frown b = (a_1,\cdots,a_{\ell(a)},b_1,\cdots b_{\ell(b)})$
  denotes the concatenation of phrases, and that for convenience,
  a single sequence element, $a_i$, may be treated as sequence of one term, $(a_i)$.
}

We manually applied the following criteria for constructing both food and exercise
phrase lists.
For a phrase to be included, it had to be a phrase that used the food
or physical activity word(s) in a way that pertained to eating or
physical activity; we excluded phrases that were part of hashtags,
Twitter user names, song lyrics, or names of organizations or
businesses, and phrases that appeared four or fewer times were not
included.
Misspellings and alternate spellings were included if we happened upon
them (for example, ``mash potatoes'' instead of ``mashed potatoes''),
but we did not go out of our way to search for them.
We queried questionable phrases to be sure that the majority of their
uses were referring to the item of interest.
Because we were building up from a small list, some specific versions
of foods were included while more general forms were not.  For
example, because we built phrases up from ``strawberry,'' ``strawberry
jam'' was included while we did not conduct a larger search for ``jam''.
In another example, in building phrases up from ``bacon,'' ``bacon
wrapped dates'' turned up so we included those dates but did not
conduct a larger search for all possible ``dates''.
(Note: We removed the physical activities category `sexual activity'
from the study because the task of determining meaning and context was
too difficult.)

We searched for phrases containing the physical activities in multiple
tenses in order to capture as much information as possible.
For example, for the activity type \textit{shoveling snow}, we
searched for the forms of shovel, shovel\textit{ing}, and
shovel\textit{ed}.
Tweets were initially converted to all lowercase text, so we were
assured that we were not missing data due to capitalization.
To match each food phrase with its closest caloric data, we found the
most closely corresponding food from the USDA National Nutrient
Database, counting all vegetables and fruits in their raw form unless
the phrase indicated otherwise.
Similarly, we entered meats as roasted or cooked with dry heat, not fried, unless the phrase
indicated otherwise or there was no homemade option.
We used the nutrition content of homemade versions of foods (for
example, baked goods) rather than store-bought foods unless the phrase
indicated otherwise.
Our approach, while systematic, was not exhaustive, nor is it the only
way of taking on this challenge; there are certainly other methods
that we expect to yield similar results.

Finally, we lemmatized the food phrases by their code in the USDA
National Nutrient Database.  If there were food phrases that were more
general in each set of phrases that held the same code, we used the
more general phrase as the lemma.

We lemmatized the activity phrases by their METs and activity
category.  Activity categories were largely the same as listed in the
Compendium with slight changes due to items in Compendium being listed
in a Miscellaneous category, etc. This yielded instances of physical
activity phrases that were in the same activity category but were very
different with the same METs being included in the same lemma.  From
this level of lemmatization, we then used our best judgement to break
these lemmas down further until proper phrases were included in each
lemma.

\acknowledgments
We thank Slack.com
and the Vermont Advanced Computing Core for 
greatly facilitating our work.
\revtexonly{
PSD was supported by NSF CAREER Grant No. 0846668,
and CMD and PSD were supported by NSF BIGDATA Grant No. 1447634.
}

\clearpage

 \newwrite\tempfile
 \immediate\openout\tempfile=startsupp.txt
 \immediate\write\tempfile{\thepage}
 \immediate\closeout\tempfile
 
 \setcounter{page}{1}
 \renewcommand{\thepage}{S\arabic{page}}
 \renewcommand{\thefigure}{S\arabic{figure}}
 \renewcommand{\thetable}{S\arabic{table}}
 \setcounter{figure}{0}
 \setcounter{table}{0}

\begin{figure*}[tp!]
    \begin{center}
      \includegraphics[width=\textwidth]{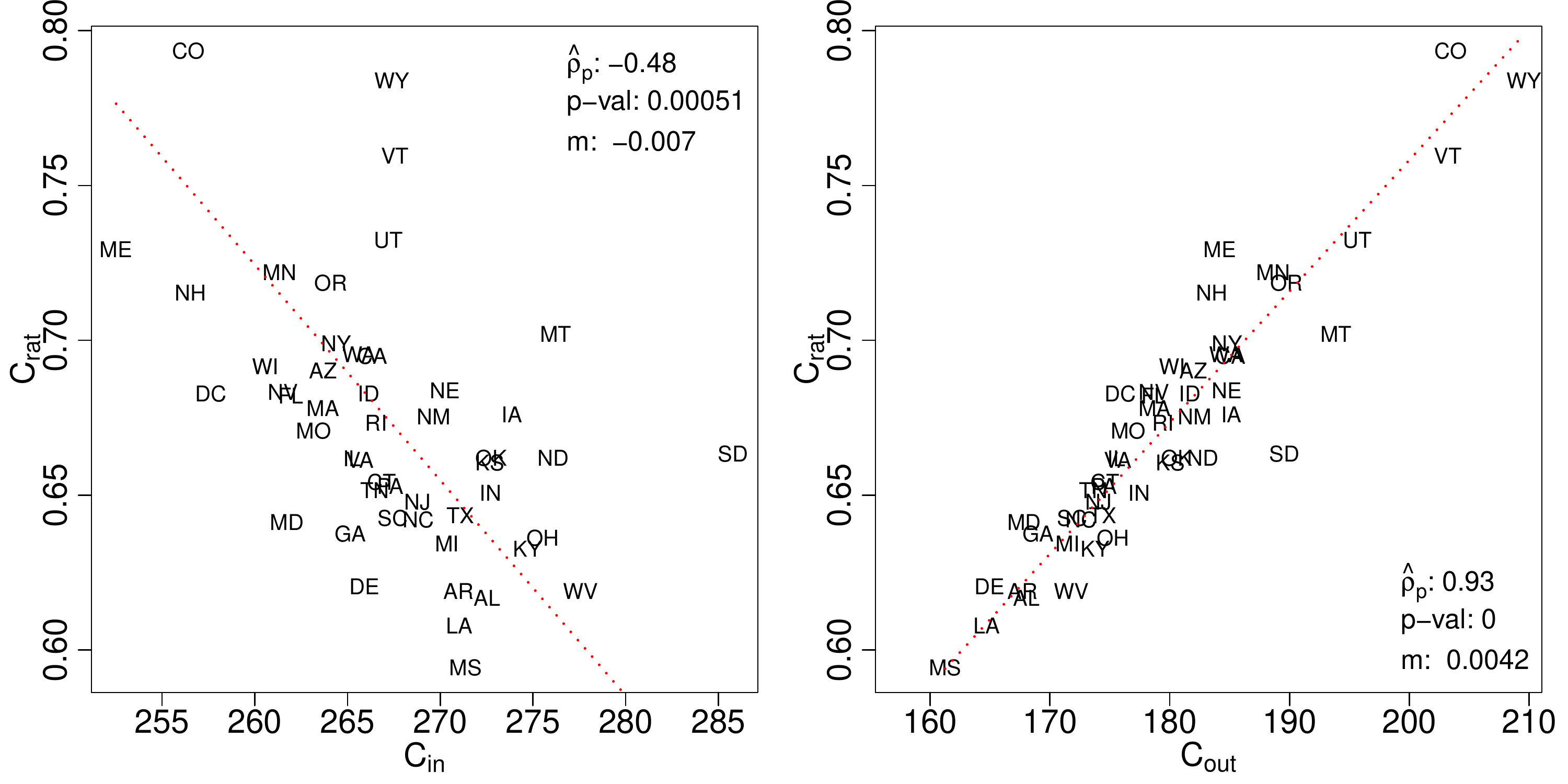}
    \end{center}
    \caption{
      Plots for the contiguous US 
      showing the relationships
      $\calrat$ versus $\calin$ (left),
      and
      $\calrat$ versus $\calout$ (right).
      With its larger range, caloric output $\calout$ 
      is more tightly coupled with the ratio $\calrat$.
    }
    \label{fig:fluxwell.supp.scatterplots}
\end{figure*}

\begin{figure*}[tp!]
    \begin{center}
      \includegraphics[width=\textwidth]{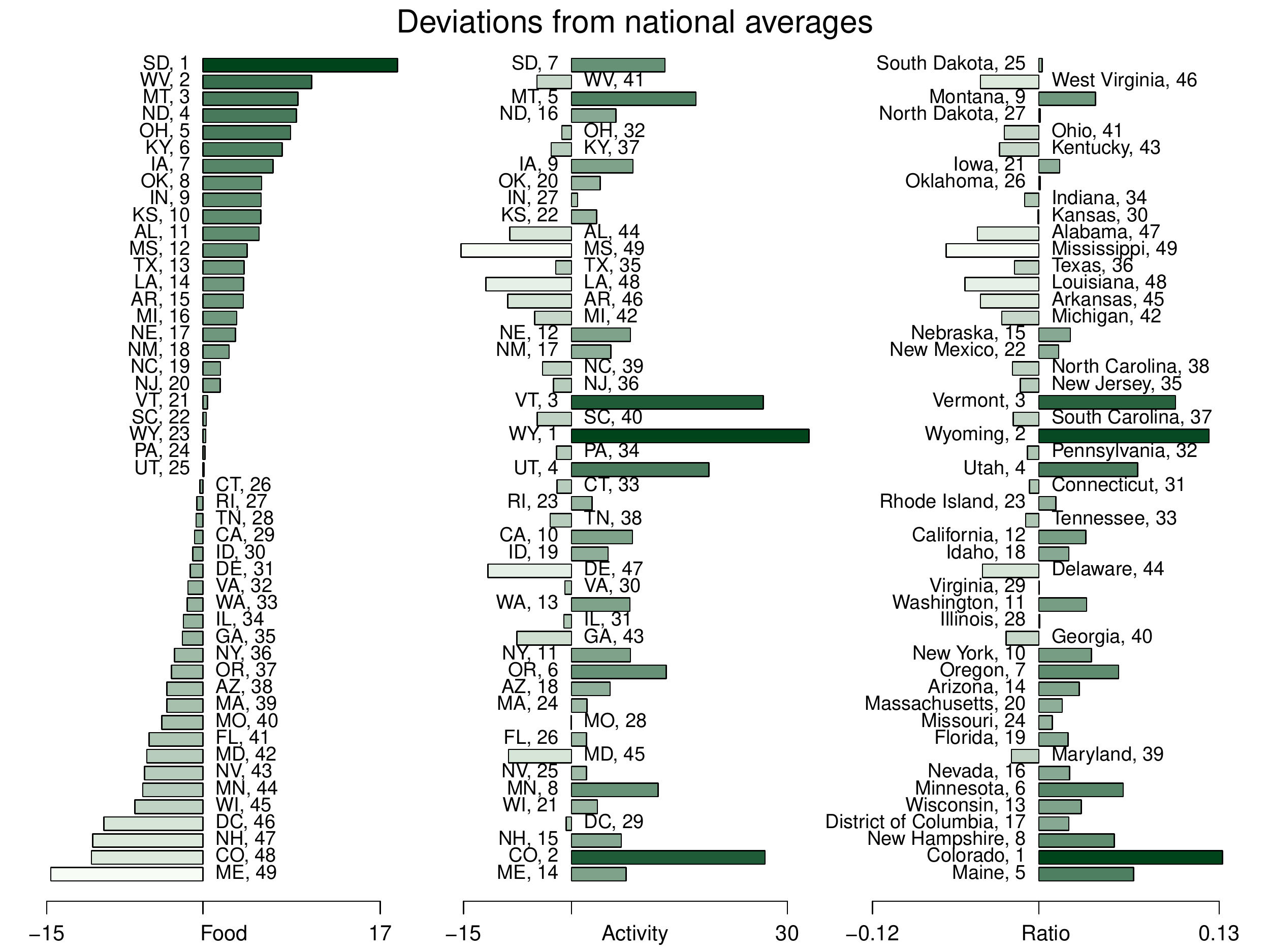}
    \end{center}
    \caption{
      Histograms as per Fig.~\ref{fig:fluxwell.histograms} with
      states sorted by food rank.
      The bar colors correspond those used in for the choropleth maps in
      Figs.~\ref{fig:fluxwell.maps-dominantmodes},
      \ref{fig:fluxwell.maps-diffs},
      and 
      \ref{fig:fluxwell.maps-calrat}.
    }
    \label{fig:fluxwell.supp.histograms-food}
\end{figure*}

\begin{figure*}[tp!]
    \begin{center}
      \includegraphics[width=\textwidth]{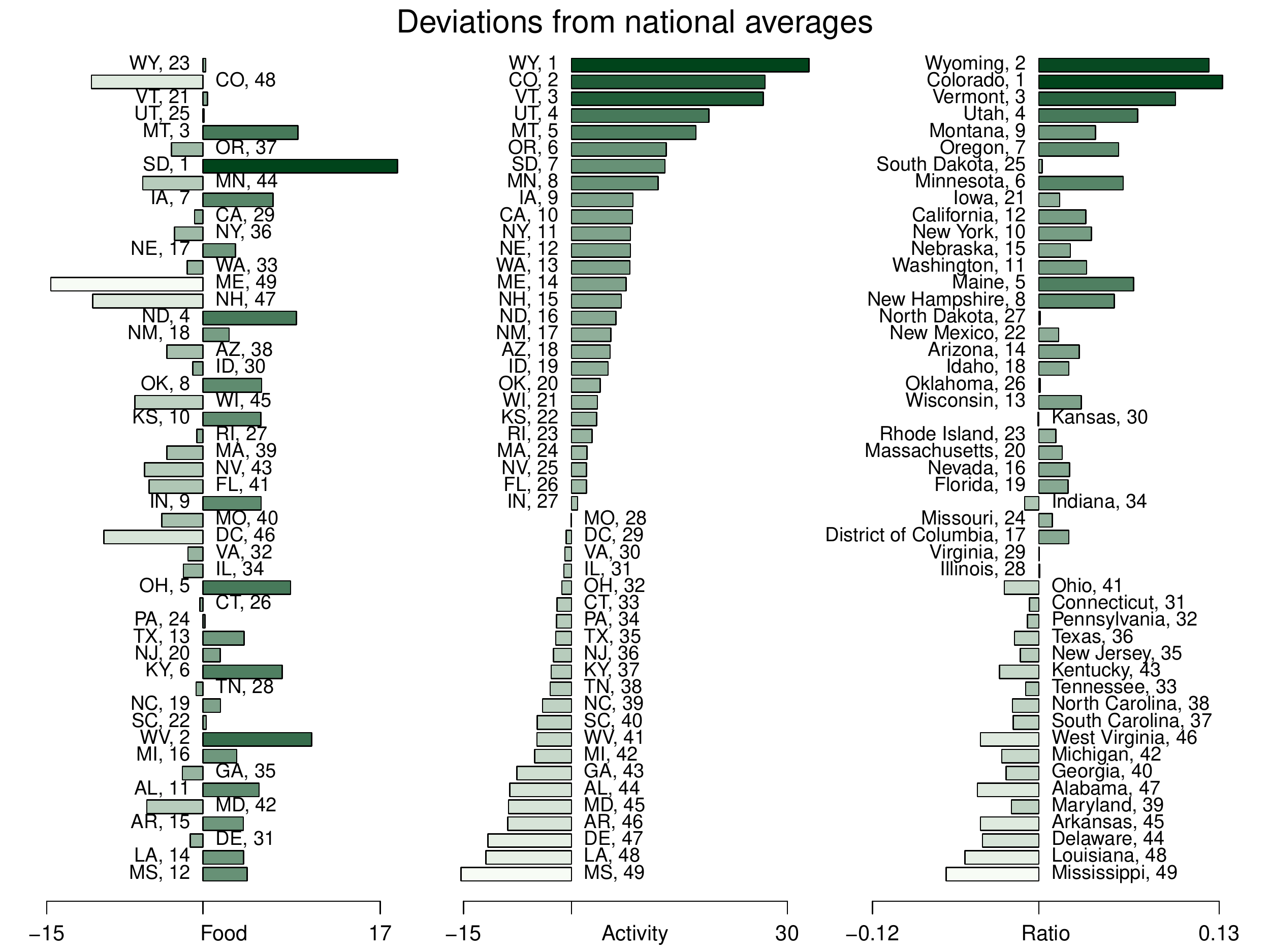}
    \end{center}
    \caption{
      Histograms as per Fig.~\ref{fig:fluxwell.histograms} with
      states sorted by activity rank.
      The bar colors correspond those used in for the choropleth maps in
      Figs.~\ref{fig:fluxwell.maps-dominantmodes},
      \ref{fig:fluxwell.maps-diffs},
      and 
      \ref{fig:fluxwell.maps-calrat}.
    }
    \label{fig:fluxwell.supp.histograms-activities}
\end{figure*}

\begin{figure*}[tp!]
    \begin{center}
      \includegraphics[width=\textwidth]{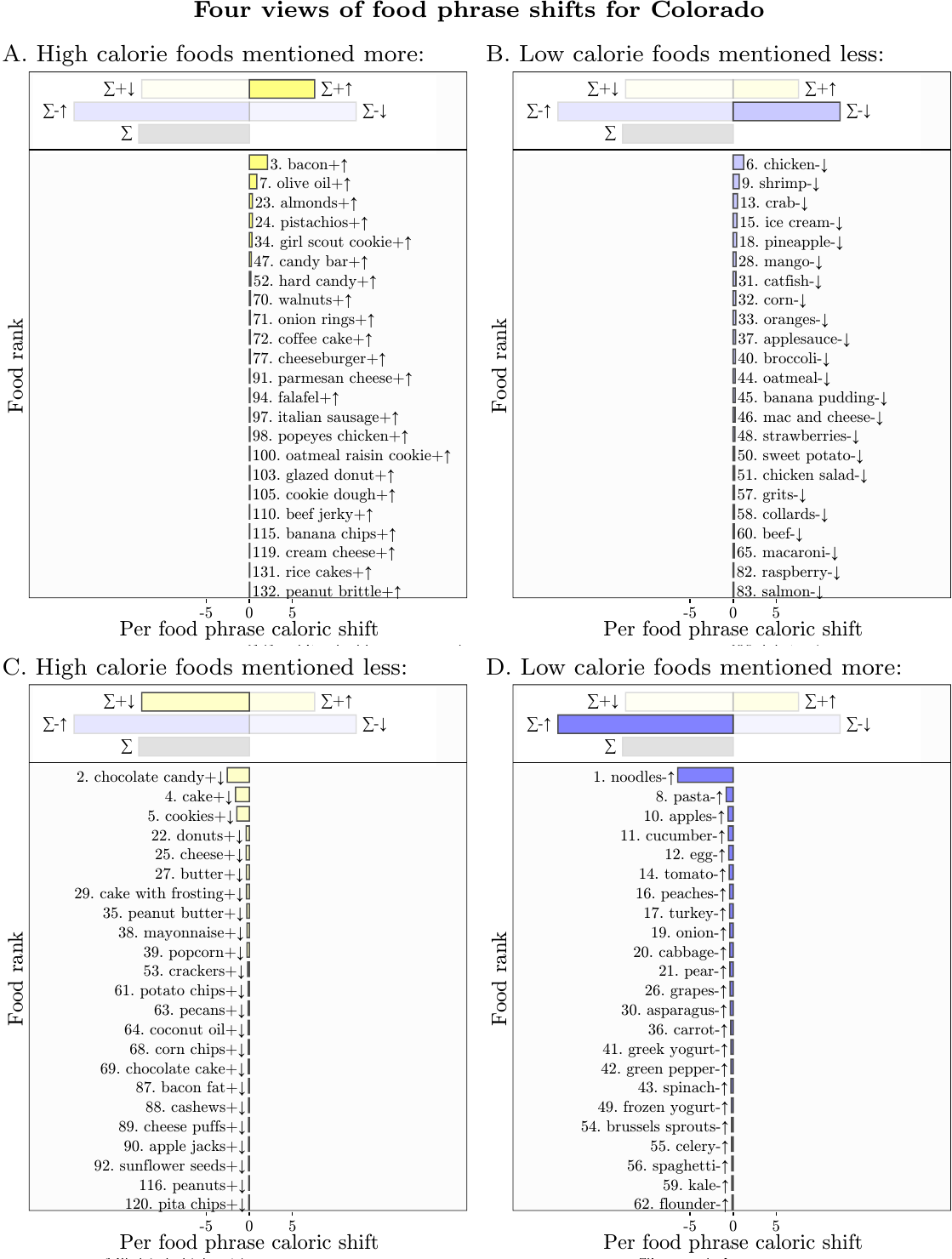}
    \end{center}
    \caption{
      Food phrase shifts for Colorado, 
      broken down 
      into the four ways phrases
      may contribute to a shift.
      See Fig.~\ref{fig:fluxwell.wordshiftexample}A for the combined shift.
      \revtexlatexswitch{See Subsec.~Phrase~Shifts in
        Sec.~Analysis~and~Results for an explanation of phrase shifts.}{
        See Phrase~Shifts in the Analysis~and~Results section for an
        explanation of phrase shifts.}
    }
    \label{fig:fluxwell.supp.colorado-food}
\end{figure*}

\begin{figure*}[tp!]
    \begin{center}
      \includegraphics[width=\textwidth]{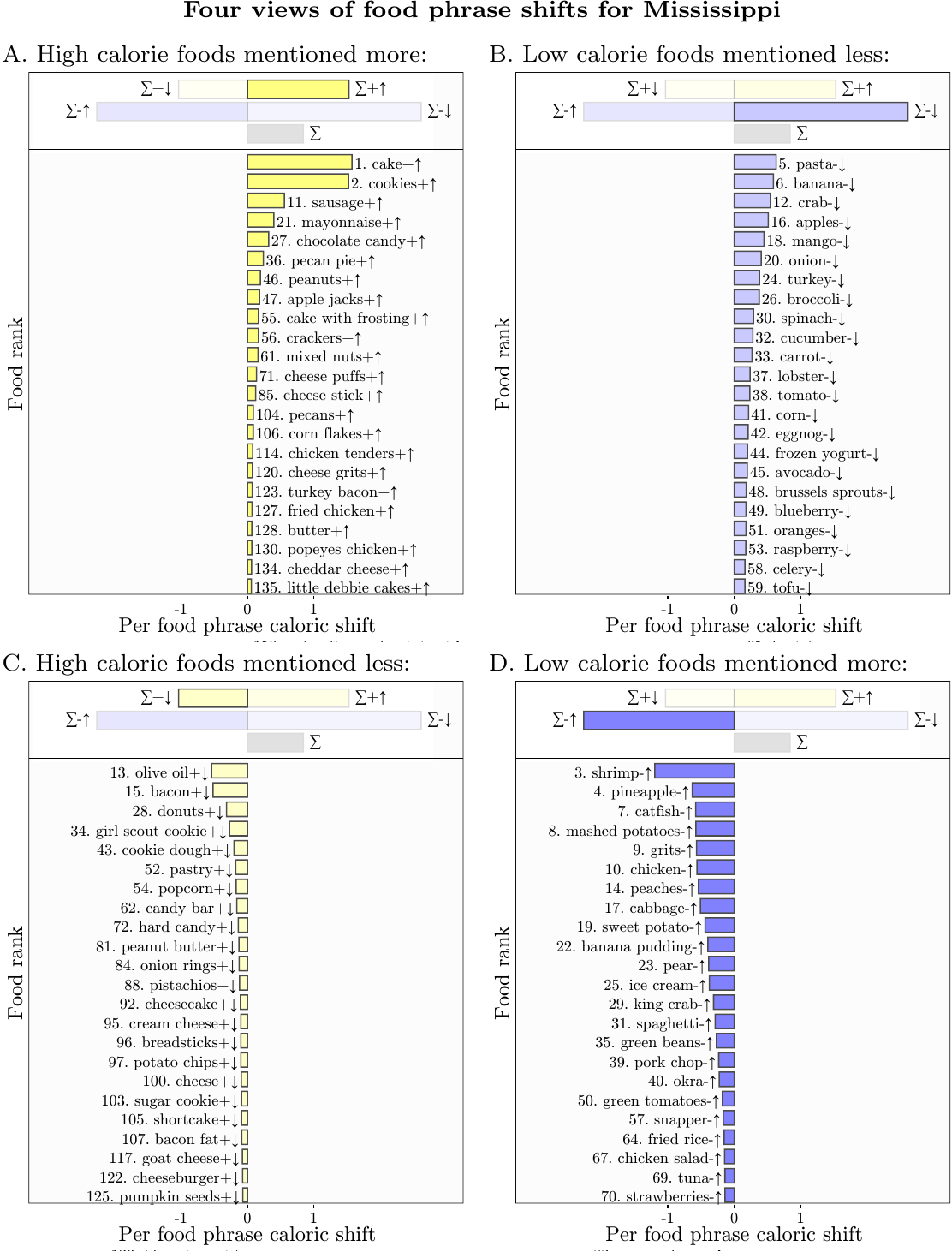}
    \end{center}
    \caption{
      Food phrase shifts for Mississippi, 
      broken down 
      into the four ways phrases
      may contribute to a shift.
      See Fig.~\ref{fig:fluxwell.wordshiftexample}B for the combined shift.
      \revtexlatexswitch{See Subsec.~Phrase~Shifts in
        Sec.~Analysis~and~Results for an explanation of phrase shifts.}{
        See Phrase~Shifts in the Analysis~and~Results section for an
        explanation of phrase shifts.}
    }
    \label{fig:fluxwell.supp.mississippi-food}
\end{figure*}

\begin{figure*}[tp!]
    \begin{center}
      \includegraphics[width=\textwidth]{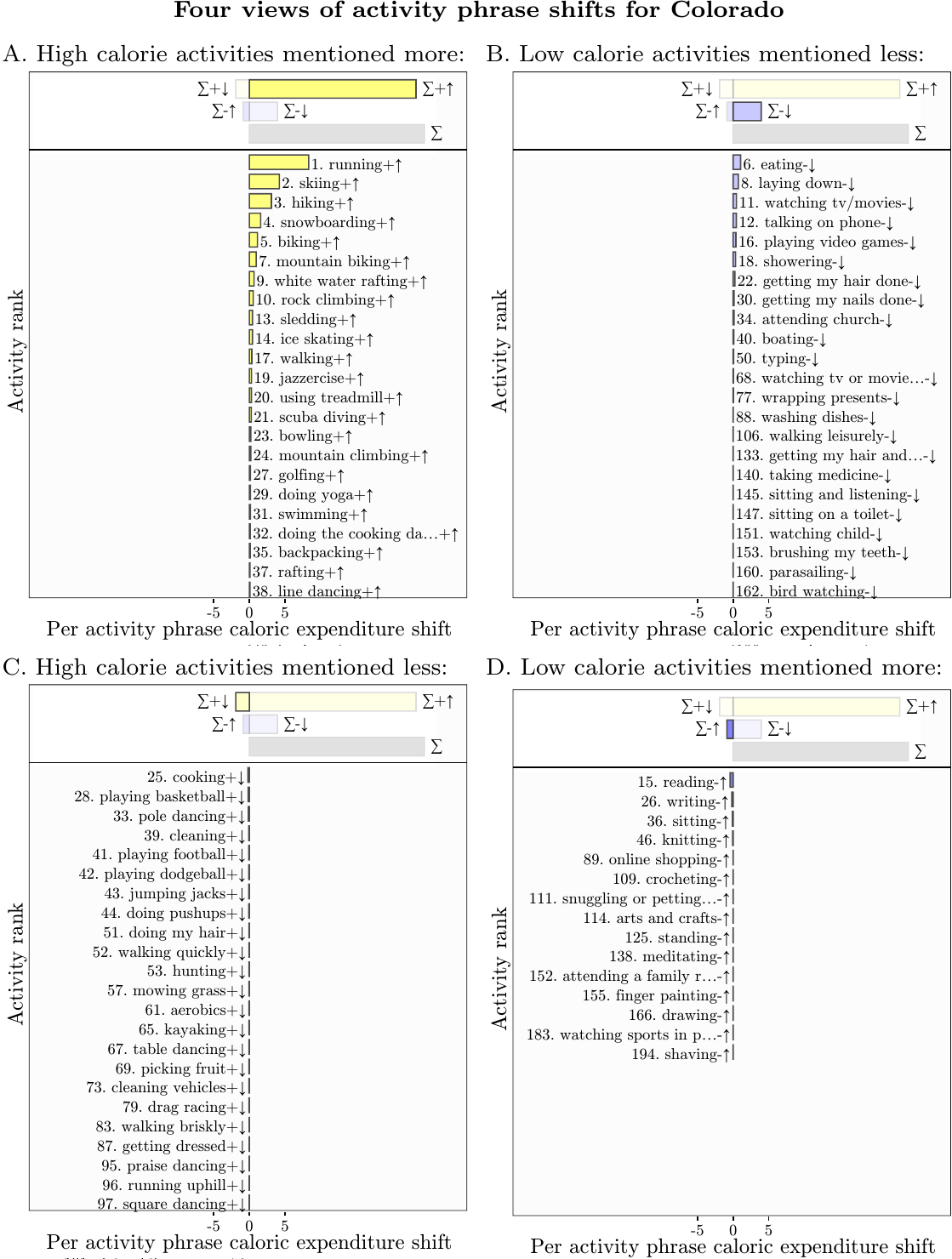}
    \end{center}
    \caption{
      Activity phrase shifts for Colorado, 
      broken down 
      into the four ways phrases
      may contribute to a shift.
      See Fig.~\ref{fig:fluxwell.wordshiftexample}C for the combined shift.
      \revtexlatexswitch{See Subsec.~Phrase~Shifts in
        Sec.~Analysis~and~Results for an explanation of phrase shifts.}{
        See Phrase~Shifts in the Analysis~and~Results section for an
        explanation of phrase shifts.}
    }
    \label{fig:fluxwell.supp.colorado-activity}
\end{figure*}

\begin{figure*}[tp!]
    \begin{center}
      \includegraphics[width=\textwidth]{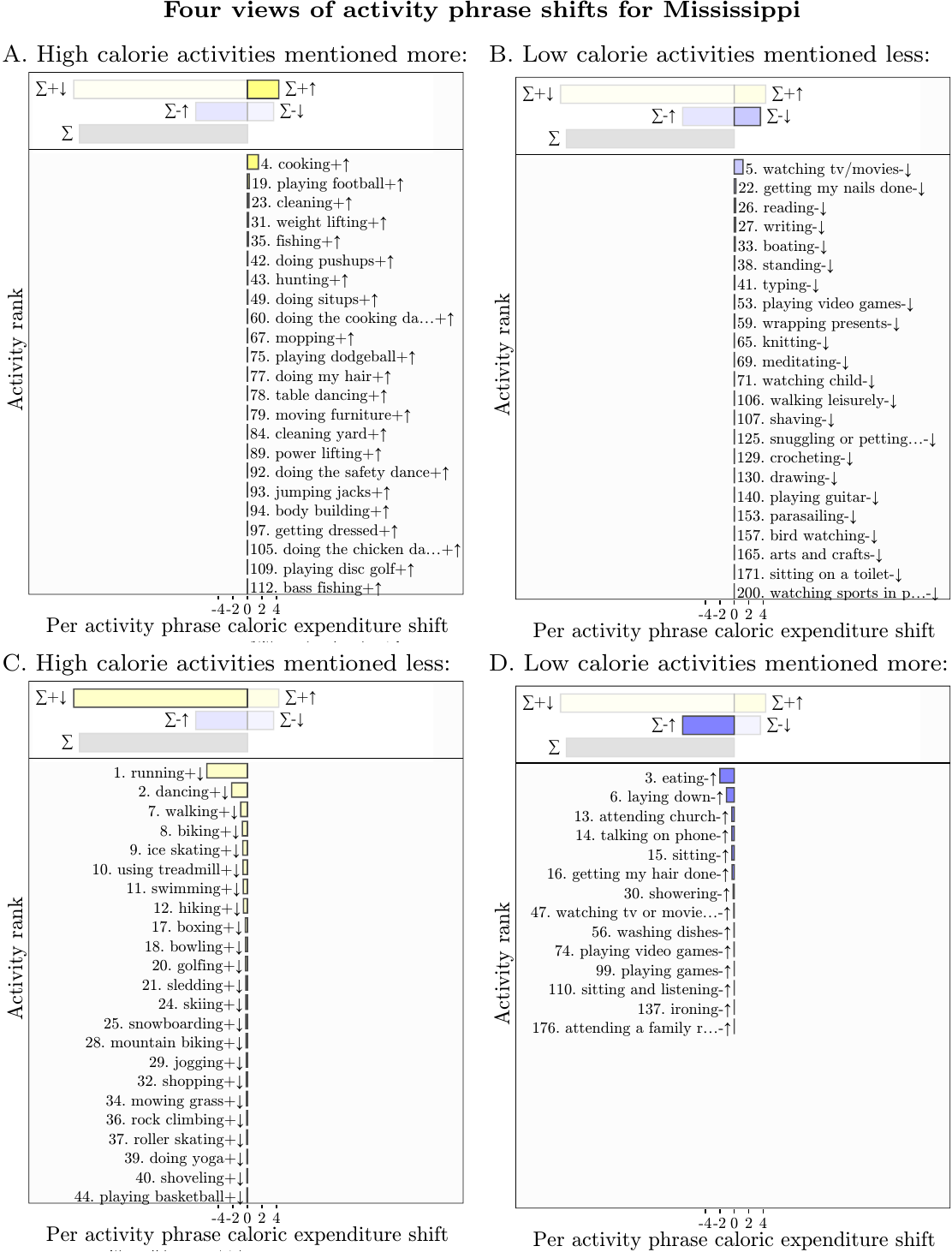}
    \end{center}
    \caption{
      Activity phrase shifts for Mississippi,
      broken down 
      into the four ways phrases
      may contribute to a shift.
      See Fig.~\ref{fig:fluxwell.wordshiftexample}D for the combined shift.
      \revtexlatexswitch{See Subsec.~Phrase~Shifts in
        Sec.~Analysis~and~Results for an explanation of phrase shifts.}{
        See Phrase~Shifts in the Analysis~and~Results section for an
        explanation of phrase shifts.}
    }
    \label{fig:fluxwell.supp.mississippi-activity}
\end{figure*}

\begin{table*}
  \revtexlatexswitch{  \plosoneonly{\begin{adjustwidth}{-4cm}{}}
  \footnotesize
  \begin{tabular}{|l|c|c|c|c|c|c|}
    \hline \multicolumn{1}{|c|}{\textbf{Health and/or well-being quantity}}
    & \begin{tabular}{@{}c@{}} $\rhospearman$ for \\ $\calrat$ 
      \end{tabular} 
    & \textbf{$q$-val}
    & \begin{tabular}{@{}c@{}}$\rhospearman$ for \\ $\calin$ 
      \end{tabular} & \textbf{$q$-val} & 
    \begin{tabular}{@{}c@{}}$\rhospearman$ for \\ 
      $\calout$ \end{tabular} & 
    \textbf{$q$-val} \\\hline
    \textbf{1.}  \% no physical activity in past 30 days \cite{america} & -0.78 & $3.07\times 10^{-09}$ & 0.58 & $4.91\times 10^{-05}$ & -0.66 & $1.59\times 10^{-06}$ \\\hline
    \textbf{2.}  \% have been physically active in past 30 days \cite{america} & 0.78 & $3.07\times 10^{-09}$ & -0.58 & $5.50\times 10^{-05}$ & 0.67 & $1.31\times 10^{-06}$ \\\hline
    \textbf{3.}  \% high blood pressure \cite{america} & -0.77 & $3.07\times 10^{-09}$ & 0.39 & $1.16\times 10^{-02}$ & -0.78 & $3.07\times 10^{-09}$ \\\hline
    \textbf{4.}  Heart disease death rate \cite{kff} & -0.75 & $1.02\times 10^{-08}$ & 0.38 & $1.24\times 10^{-02}$ & -0.73 & $2.07\times 10^{-08}$ \\\hline
    \textbf{5.}  Adult diabetes rate \cite{cdc} & -0.74 & $1.17\times 10^{-08}$ & 0.34 & $2.77\times 10^{-02}$ & -0.77 & $3.07\times 10^{-09}$ \\\hline
    \textbf{6.}  CNBC quality of life ranking \cite{cnbc} & -0.74 & $1.87\times 10^{-08}$ & 0.33 & $3.22\times 10^{-02}$ & -0.77 & $3.60\times 10^{-09}$ \\\hline
    \textbf{7.}  \% adult overweight/obesity \cite{kff} & -0.71 & $1.33\times 10^{-07}$ & 0.53 & $3.14\times 10^{-04}$ & -0.59 & $3.56\times 10^{-05}$ \\\hline
    \textbf{8.}  Gallup Wellbeing score \cite{gallup} & 0.7 & $3.17\times 10^{-07}$ & -0.33 & $3.38\times 10^{-02}$ & 0.73 & $4.35\times 10^{-08}$ \\\hline
    \textbf{9.}  \% adult obesity \cite{cdc} & -0.69 & $3.10\times 10^{-07}$ & 0.52 & $4.11\times 10^{-04}$ & -0.59 & $3.56\times 10^{-05}$ \\\hline
    \textbf{10.}  America's Health Rankings, overall \cite{america} & -0.69 & $1.31\times 10^{-06}$ & 0.4 & $9.14\times 10^{-03}$ & -0.67 & $2.65\times 10^{-06}$ \\\hline
    \textbf{11.}  Life expectancy at birth \cite{kff}  & 0.67 & $7.92\times 10^{-07}$ & -0.36 & $1.59\times 10^{-02}$ & 0.65 & $2.58\times 10^{-06}$ \\\hline
    \textbf{12.}  \% child overweight/obesity \cite{kff} & -0.65 & $2.58\times 10^{-06}$ & 0.34 & $2.82\times 10^{-02}$ & -0.64 & $3.06\times 10^{-06}$ \\\hline
    \textbf{13.}  \% who eat fruit less than once a day \cite{produce} & -0.65 & $2.58\times 10^{-06}$ & 0.57 & $7.45\times 10^{-05}$ & -0.51 & $5.89\times 10^{-04}$ \\\hline
    \textbf{14.}  \% who eat vegetables less than once a day \cite{produce} & -0.61 & $1.32\times 10^{-05}$ & 0.53 & $3.14\times 10^{-04}$ & -0.46 & $1.72\times 10^{-03}$ \\\hline
    \textbf{15.}  Median daily intake of fruits \cite{produce} & 0.59 & $3.56\times 10^{-05}$ & -0.59 & $3.56\times 10^{-05}$ & 0.41 & $5.73\times 10^{-03}$ \\\hline
    \textbf{16.}  Smoking rate \cite{kff}  & -0.59 & $3.81\times 10^{-05}$ & 0.47 & $1.60\times 10^{-03}$ & -0.48 & $1.24\times 10^{-03}$ \\\hline
    \textbf{17.}  Median daily intake of vegetables \cite{produce} & 0.5 & $7.25\times 10^{-04}$ & -0.56 & $1.03\times 10^{-04}$ & 0.31 & $4.09\times 10^{-02}$ \\\hline
    \textbf{18.}  Median household income \cite{kff} & 0.48 & $1.37\times 10^{-03}$ & -0.5 & $8.58\times 10^{-04}$ & 0.4 & $9.07\times 10^{-03}$ \\\hline
    \textbf{19.}  \% high cholesterol \cite{america} & -0.48 & $1.26\times 10^{-03}$ & 0.24 & $1.16\times 10^{-01}$ & -0.48 & $1.05\times 10^{-03}$ \\\hline
    \textbf{20.}  Colorectal cancer rate \cite{cdc} & -0.47 & $1.72\times 10^{-03}$ & 0.56 & $1.37\times 10^{-04}$ & -0.27 & $8.35\times 10^{-02}$ \\\hline
    \textbf{21.}  Brain health ranking \cite{brain} (lower is better) & -0.46 & $1.95\times 10^{-03}$ & 0.55 & $1.74\times 10^{-04}$ & -0.29 & $5.43\times 10^{-02}$ \\\hline
    \textbf{22.}  US Census Gini index score \cite{gini} (lower is better) & -0.44 & $3.60\times 10^{-03}$ & 0.11 & $5.12\times 10^{-01}$ & -0.5 & $6.22\times 10^{-04}$ \\\hline
    \textbf{23.}  \% with bachelor's degree or higher \cite{census} & 0.42 & $4.86\times 10^{-03}$ & -0.43 & $4.21\times 10^{-03}$ & 0.33 & $2.82\times 10^{-02}$ \\\hline
    \textbf{24.}  Avg \# poor mental health days, past 30 days \cite{america} & -0.39 & $9.87\times 10^{-03}$ & 0.1 & $5.31\times 10^{-01}$ & -0.48 & $1.23\times 10^{-03}$ \\\hline
    \textbf{25.}  Neuroticism Big Five personality trait \cite{rentfrow} & -0.37 & $1.33\times 10^{-02}$ & 0.23 & $1.35\times 10^{-01}$ & -0.37 & $1.42\times 10^{-02}$ \\\hline
    \textbf{26.}  Binge drinking rate \cite{america} & 0.34 & $2.91\times 10^{-02}$ & -0.12 & $4.88\times 10^{-01}$ & 0.41 & $6.23\times 10^{-03}$ \\\hline
    \textbf{27.}  Farmers markets per 100,000 in pop. \cite{produce} & 0.33 & $2.96\times 10^{-02}$ & -0.01 & $9.59\times 10^{-01}$ & 0.42 & $5.41\times 10^{-03}$ \\\hline
    \textbf{28.}  Extraversion Big Five personality trait \cite{rentfrow} & -0.33 & $2.83\times 10^{-02}$ & 0.13 & $4.13\times 10^{-01}$ & -0.29 & $5.36\times 10^{-02}$ \\\hline
    \textbf{29.}  Avg \# poor physical health days, past 30 days \cite{america} & -0.32 & $3.81\times 10^{-02}$ & 0.16 & $3.32\times 10^{-01}$ & -0.38 & $1.16\times 10^{-02}$ \\\hline
    \textbf{30.}  Strolling of the Heifers locavore score (lower is better) \cite{stroll} & -0.31 & $4.59\times 10^{-02}$ & -0.16 & $3.32\times 10^{-01}$ & -0.45 & $3.16\times 10^{-03}$ \\\hline
    \textbf{31.}  \% schools offering fruit/veg at celebrations \cite{produce} & 0.25 & $1.16\times 10^{-01}$ & -0.38 & $1.36\times 10^{-02}$ & 0.05 & $7.75\times 10^{-01}$ \\\hline
    \textbf{32.}  Openness Big Five personality trait \cite{rentfrow} & 0.23 & $1.31\times 10^{-01}$ & -0.42 & $5.43\times 10^{-03}$ & 0.04 & $7.95\times 10^{-01}$ \\\hline
    \textbf{33.}  \% cropland harvested for fruits/veg \cite{produce} & 0.18 & $2.53\times 10^{-01}$ & -0.53 & $2.90\times 10^{-04}$ & -0.04 & $7.95\times 10^{-01}$ \\\hline
    \textbf{34.}  Conscientiousness Big Five personality trait \cite{rentfrow} & -0.1 & $5.31\times 10^{-01}$ & 0.14 & $3.97\times 10^{-01}$ & -0.05 & $7.78\times 10^{-01}$ \\\hline
    \textbf{35.}  \% census tracts, healthy food retailer within 1/2 mile \cite{produce} & -0.06 & $7.47\times 10^{-01}$ & -0.39 & $1.09\times 10^{-02}$ & -0.24 & $1.28\times 10^{-01}$ \\\hline
    \textbf{36.}  George Mason overall freedom ranking \cite{freedom} (lower is freer) & -0.02 & $8.90\times 10^{-01}$ & -0.05 & $7.73\times 10^{-01}$ & -0.1 & $5.58\times 10^{-01}$ \\\hline
    \textbf{37.}  Agreeableness Big Five personality trait \cite{rentfrow} & 0 & $9.95\times 10^{-01}$ & 0.24 & $1.26\times 10^{-01}$ & 0.08 & $6.41\times 10^{-01}$ 
    \\\hline
  \end{tabular}
}{}
  \caption{
        Identical to Tab.~\ref{tab:fluxwell.corrTable} but with liquids included.
    Spearman correlation coefficients, $\rhospearman$, and Benjamini-Hochberg $q$-values for
    caloric input $\calin$,
    caloric output $\calout$, 
    and caloric ratio $\calrat = \calout/\calin$
    and
    demographic data related to food and physical activity, Big Five
    personality traits~\cite{rentfrow}, health and well-being rankings by state, and
    socioeconomic status, correlated, ordered from strongest to weakest Spearman
    correlations with caloric ratio.
  }
  \label{tab:fluxwell.supp.corrTableLIQ}
  \plosoneonly{\end{adjustwidth}}
\end{table*}

\begin{table*}
  \revtexlatexswitch{  \plosoneonly{\begin{adjustwidth}{-4cm}{}}
  \footnotesize
  \begin{tabular}{|l|c|c|c|c|c|c|}
    \hline \multicolumn{1}{|c|}{\textbf{Health and/or well-being quantity}}
    & \begin{tabular}{@{}c@{}} $\rhospearman$ for \\ $\caldiff$ 
      \end{tabular} 
    & \textbf{$q$-val}
    & \begin{tabular}{@{}c@{}}$\rhospearman$ for \\ $\calin$ 
      \end{tabular} & \textbf{$q$-val} & 
    \begin{tabular}{@{}c@{}}$\rhospearman$ for \\ 
      $\calout$ \end{tabular} & 
    \textbf{$q$-val} \\\hline
    \textbf{1.}  \% no physical activity in past 30 days \cite{america} & -0.79 & $1.77\times 10^{-09}$ & 0.58 & $5.67\times 10^{-05}$ & -0.66 & $1.51\times 10^{-06}$ \\\hline
    \textbf{2.}  \% have been physically active in past 30 days \cite{america} & 0.79 & $1.77\times 10^{-09}$ & -0.57 & $6.53\times 10^{-05}$ & 0.67 & $1.24\times 10^{-06}$ \\\hline
    \textbf{3.}  \% high blood pressure \cite{america} & -0.78 & $2.72\times 10^{-09}$ & 0.32 & $4.05\times 10^{-02}$ & -0.78 & $2.72\times 10^{-09}$ \\\hline
    \textbf{4.}  Adult diabetes rate \cite{cdc} & -0.76 & $5.26\times 10^{-09}$ & 0.29 & $6.16\times 10^{-02}$ & -0.77 & $2.73\times 10^{-09}$ \\\hline
    \textbf{5.}  CNBC quality of life ranking \cite{cnbc} & -0.75 & $8.07\times 10^{-09}$ & 0.28 & $7.34\times 10^{-02}$ & -0.77 & $3.60\times 10^{-09}$ \\\hline
    \textbf{6.}  \% adult overweight/obesity \cite{kff} & -0.73 & $2.40\times 10^{-08}$ & 0.55 & $1.41\times 10^{-04}$ & -0.59 & $3.07\times 10^{-05}$ \\\hline
    \textbf{7.}  Heart disease death rate \cite{kff} & -0.73 & $2.07\times 10^{-08}$ & 0.34 & $2.82\times 10^{-02}$ & -0.73 & $2.07\times 10^{-08}$ \\\hline
    \textbf{8.}  Gallup Wellbeing score \cite{gallup} & 0.73 & $3.83\times 10^{-08}$ & -0.31 & $4.43\times 10^{-02}$ & 0.73 & $3.70\times 10^{-08}$ \\\hline
    \textbf{9.}  \% adult obesity \cite{cdc} & -0.72 & $3.70\times 10^{-08}$ & 0.53 & $2.26\times 10^{-04}$ & -0.59 & $2.94\times 10^{-05}$ \\\hline
    \textbf{10.}  America's Health Rankings, overall \cite{america} & -0.72 & $3.93\times 10^{-07}$ & 0.43 & $4.74\times 10^{-03}$ & -0.67 & $2.77\times 10^{-06}$ \\\hline
    \textbf{11.}  Life expectancy at birth \cite{kff}  & 0.68 & $4.27\times 10^{-07}$ & -0.4 & $6.91\times 10^{-03}$ & 0.65 & $2.64\times 10^{-06}$ \\\hline
    \textbf{12.}  \% who eat fruit less than once a day \cite{produce} & -0.67 & $9.44\times 10^{-07}$ & 0.61 & $1.38\times 10^{-05}$ & -0.51 & $5.23\times 10^{-04}$ \\\hline
    \textbf{13.}  \% child overweight/obesity \cite{kff} & -0.64 & $3.03\times 10^{-06}$ & 0.27 & $7.55\times 10^{-02}$ & -0.64 & $3.06\times 10^{-06}$ \\\hline
    \textbf{14.}  \% who eat vegetables less than once a day \cite{produce} & -0.61 & $1.38\times 10^{-05}$ & 0.51 & $5.21\times 10^{-04}$ & -0.46 & $1.57\times 10^{-03}$ \\\hline
    \textbf{15.}  Median daily intake of fruits \cite{produce} & 0.6 & $1.68\times 10^{-05}$ & -0.62 & $8.33\times 10^{-06}$ & 0.41 & $5.44\times 10^{-03}$ \\\hline
    \textbf{16.}  Smoking rate \cite{kff}  & -0.6 & $2.14\times 10^{-05}$ & 0.51 & $5.19\times 10^{-04}$ & -0.48 & $1.08\times 10^{-03}$ \\\hline
    \textbf{17.}  Median household income \cite{kff} & 0.51 & $5.19\times 10^{-04}$ & -0.53 & $3.27\times 10^{-04}$ & 0.4 & $8.38\times 10^{-03}$ \\\hline
    \textbf{18.}  Median daily intake of vegetables \cite{produce} & 0.5 & $5.72\times 10^{-04}$ & -0.56 & $7.44\times 10^{-05}$ & 0.31 & $4.36\times 10^{-02}$ \\\hline
    \textbf{19.}  Brain health ranking \cite{brain} (lower is better) & -0.5 & $7.50\times 10^{-04}$ & 0.62 & $1.38\times 10^{-05}$ & -0.29 & $5.70\times 10^{-02}$ \\\hline
    \textbf{20.}  \% high cholesterol \cite{america} & -0.49 & $7.88\times 10^{-04}$ & 0.23 & $1.45\times 10^{-01}$ & -0.48 & $9.05\times 10^{-04}$ \\\hline
    \textbf{21.}  \% with bachelor's degree or higher \cite{census} & 0.47 & $1.48\times 10^{-03}$ & -0.54 & $1.66\times 10^{-04}$ & 0.33 & $2.82\times 10^{-02}$ \\\hline
    \textbf{22.}  Colorectal cancer rate \cite{cdc} & -0.44 & $3.82\times 10^{-03}$ & 0.53 & $3.59\times 10^{-04}$ & -0.27 & $8.25\times 10^{-02}$ \\\hline
    \textbf{23.}  US Census Gini index score \cite{gini} (lower is better) & -0.42 & $4.99\times 10^{-03}$ & -0.03 & $8.45\times 10^{-01}$ & -0.5 & $5.55\times 10^{-04}$ \\\hline
    \textbf{24.}  Avg \# poor mental health days, past 30 days \cite{america} & -0.42 & $5.44\times 10^{-03}$ & 0.12 & $4.75\times 10^{-01}$ & -0.48 & $1.06\times 10^{-03}$ \\\hline
    \textbf{25.}  Neuroticism Big Five personality trait \cite{rentfrow} & -0.38 & $1.13\times 10^{-02}$ & 0.2 & $2.03\times 10^{-01}$ & -0.37 & $1.42\times 10^{-02}$ \\\hline
    \textbf{26.}  Binge drinking rate \cite{america} & 0.38 & $1.32\times 10^{-02}$ & -0.15 & $3.56\times 10^{-01}$ & 0.41 & $5.84\times 10^{-03}$ \\\hline
    \textbf{27.}  Avg \# poor physical health days, past 30 days \cite{america} & -0.35 & $2.34\times 10^{-02}$ & 0.19 & $2.19\times 10^{-01}$ & -0.38 & $1.13\times 10^{-02}$ \\\hline
    \textbf{28.}  Farmers markets per 100,000 in pop. \cite{produce} & 0.33 & $2.82\times 10^{-02}$ & 0.06 & $7.17\times 10^{-01}$ & 0.42 & $5.05\times 10^{-03}$ \\\hline
    \textbf{29.}  Strolling of the Heifers locavore score (lower is better) \cite{stroll} & -0.29 & $6.44\times 10^{-02}$ & -0.3 & $5.41\times 10^{-02}$ & -0.45 & $2.94\times 10^{-03}$ \\\hline
    \textbf{30.}  Extraversion Big Five personality trait \cite{rentfrow} & -0.28 & $6.89\times 10^{-02}$ & 0.03 & $8.50\times 10^{-01}$ & -0.29 & $5.63\times 10^{-02}$ \\\hline
    \textbf{31.}  \% schools offering fruit/veg at celebrations \cite{produce} & 0.24 & $1.26\times 10^{-01}$ & -0.46 & $1.96\times 10^{-03}$ & 0.05 & $7.90\times 10^{-01}$ \\\hline
    \textbf{32.}  Openness Big Five personality trait \cite{rentfrow} & 0.24 & $1.26\times 10^{-01}$ & -0.5 & $6.11\times 10^{-04}$ & 0.04 & $8.10\times 10^{-01}$ \\\hline
    \textbf{33.}  \% cropland harvested for fruits/veg \cite{produce} & 0.19 & $2.35\times 10^{-01}$ & -0.62 & $1.37\times 10^{-05}$ & -0.04 & $8.10\times 10^{-01}$ \\\hline
    \textbf{34.}  Conscientiousness Big Five personality trait \cite{rentfrow} & -0.12 & $4.62\times 10^{-01}$ & 0.2 & $2.10\times 10^{-01}$ & -0.05 & $7.93\times 10^{-01}$ \\\hline
    \textbf{35.}  \% census tracts, healthy food retailer within 1/2 mile \cite{produce} & -0.02 & $8.86\times 10^{-01}$ & -0.52 & $3.68\times 10^{-04}$ & -0.24 & $1.28\times 10^{-01}$ \\\hline
    \textbf{36.}  George Mason overall freedom ranking \cite{freedom} (lower is freer) & -0.02 & $8.88\times 10^{-01}$ & -0.11 & $5.15\times 10^{-01}$ & -0.1 & $5.64\times 10^{-01}$ \\\hline
    \textbf{37.}  Agreeableness Big Five personality trait \cite{rentfrow} & -0.01 & $9.42\times 10^{-01}$ & 0.22 & $1.50\times 10^{-01}$ & 0.08 & $6.47\times 10^{-01}$
    \\\hline
  \end{tabular}
}{}
  \caption{
        Identical to Tab.~\ref{tab:fluxwell.corrTable} but using a caloric
    difference
    rather than caloric ratio.
    Spearman correlation coefficients, $\rhospearman$, and Benjamini-Hochberg $q$-values for
    caloric input $\calin$,
    caloric output $\calout$, 
    and caloric difference $\caldiff(\alpha) = \alpha \calout + (1-\alpha)\calin$
    and
    demographic data related to food and physical activity, Big Five
    personality traits~\cite{rentfrow}, health and well-being rankings by state, and
    socioeconomic status, correlated, ordered from strongest to weakest Spearman
    correlations with caloric ratio.
    We chose $\alpha$ so that the average of $\calout$ matched the
    average of $\alpha \calin$.
  }
  \label{tab:fluxwell.supp.corrTableNOLIQ_DIF}
  \plosoneonly{\end{adjustwidth}}
\end{table*}

\begin{table*}
  \revtexlatexswitch{  \plosoneonly{\begin{adjustwidth}{-4cm}{}}
  \footnotesize
  \begin{tabular}{|l|c|c|c|c|c|c|}
    \hline \multicolumn{1}{|c|}{\textbf{Health and/or well-being quantity}}
    & \begin{tabular}{@{}c@{}} $\rhospearman$ for \\ $\caldiff$ 
      \end{tabular} 
    & \textbf{$q$-val}
    & \begin{tabular}{@{}c@{}}$\rhospearman$ for \\ $\calin$ 
      \end{tabular} & \textbf{$q$-val} & 
    \begin{tabular}{@{}c@{}}$\rhospearman$ for \\ 
      $\calout$ \end{tabular} & 
    \textbf{$q$-val} \\\hline
    \textbf{1.}  \% no physical activity in past 30 days \cite{america} & -0.78 & $3.42\times 10^{-09}$ & 0.58 & $4.91\times 10^{-05}$ & -0.66 & $1.59\times 10^{-06}$ \\\hline
    \textbf{2.}  \% have been physically active in past 30 days \cite{america} & 0.78 & $3.42\times 10^{-09}$ & -0.58 & $5.50\times 10^{-05}$ & 0.67 & $1.39\times 10^{-06}$ \\\hline
    \textbf{3.}  \% high blood pressure \cite{america} & -0.77 & $3.60\times 10^{-09}$ & 0.39 & $1.16\times 10^{-02}$ & -0.78 & $3.42\times 10^{-09}$ \\\hline
    \textbf{4.}  Heart disease death rate \cite{kff} & -0.75 & $1.09\times 10^{-08}$ & 0.38 & $1.24\times 10^{-02}$ & -0.73 & $2.07\times 10^{-08}$ \\\hline
    \textbf{5.}  Adult diabetes rate \cite{cdc} & -0.74 & $1.25\times 10^{-08}$ & 0.34 & $2.77\times 10^{-02}$ & -0.77 & $3.42\times 10^{-09}$ \\\hline
    \textbf{6.}  CNBC quality of life ranking \cite{cnbc} & -0.74 & $2.07\times 10^{-08}$ & 0.33 & $3.22\times 10^{-02}$ & -0.77 & $3.60\times 10^{-09}$ \\\hline
    \textbf{7.}  \% adult overweight/obesity \cite{kff} & -0.7 & $1.48\times 10^{-07}$ & 0.53 & $3.14\times 10^{-04}$ & -0.59 & $3.56\times 10^{-05}$ \\\hline
    \textbf{8.}  Gallup Wellbeing score \cite{gallup} & 0.7 & $3.08\times 10^{-07}$ & -0.33 & $3.38\times 10^{-02}$ & 0.73 & $4.35\times 10^{-08}$ \\\hline
    \textbf{9.}  \% adult obesity \cite{cdc} & -0.69 & $3.40\times 10^{-07}$ & 0.52 & $4.11\times 10^{-04}$ & -0.59 & $3.56\times 10^{-05}$ \\\hline
    \textbf{10.}  America's Health Rankings, overall \cite{america} & -0.69 & $1.39\times 10^{-06}$ & 0.4 & $9.14\times 10^{-03}$ & -0.67 & $2.77\times 10^{-06}$ \\\hline
    \textbf{11.}  Life expectancy at birth \cite{kff}  & 0.67 & $9.05\times 10^{-07}$ & -0.36 & $1.59\times 10^{-02}$ & 0.65 & $2.67\times 10^{-06}$ \\\hline
    \textbf{12.}  \% who eat fruit less than once a day \cite{produce} & -0.65 & $2.67\times 10^{-06}$ & 0.57 & $7.45\times 10^{-05}$ & -0.51 & $5.89\times 10^{-04}$ \\\hline
    \textbf{13.}  \% child overweight/obesity \cite{kff} & -0.64 & $3.06\times 10^{-06}$ & 0.34 & $2.78\times 10^{-02}$ & -0.64 & $3.06\times 10^{-06}$ \\\hline
    \textbf{14.}  \% who eat vegetables less than once a day \cite{produce} & -0.61 & $1.54\times 10^{-05}$ & 0.53 & $3.14\times 10^{-04}$ & -0.46 & $1.69\times 10^{-03}$ \\\hline
    \textbf{15.}  Median daily intake of fruits \cite{produce} & 0.59 & $3.56\times 10^{-05}$ & -0.59 & $3.56\times 10^{-05}$ & 0.41 & $5.73\times 10^{-03}$ \\\hline
    \textbf{16.}  Smoking rate \cite{kff}  & -0.59 & $3.77\times 10^{-05}$ & 0.47 & $1.60\times 10^{-03}$ & -0.48 & $1.24\times 10^{-03}$ \\\hline
    \textbf{17.}  Median daily intake of vegetables \cite{produce} & 0.5 & $7.64\times 10^{-04}$ & -0.56 & $1.03\times 10^{-04}$ & 0.31 & $4.09\times 10^{-02}$ \\\hline
    \textbf{18.}  Median household income \cite{kff} & 0.48 & $1.38\times 10^{-03}$ & -0.5 & $8.58\times 10^{-04}$ & 0.4 & $9.07\times 10^{-03}$ \\\hline
    \textbf{19.}  \% high cholesterol \cite{america} & -0.48 & $1.28\times 10^{-03}$ & 0.24 & $1.15\times 10^{-01}$ & -0.48 & $1.05\times 10^{-03}$ \\\hline
    \textbf{20.}  Colorectal cancer rate \cite{cdc} & -0.47 & $1.68\times 10^{-03}$ & 0.56 & $1.37\times 10^{-04}$ & -0.27 & $8.35\times 10^{-02}$ \\\hline
    \textbf{21.}  Brain health ranking \cite{brain} (lower is better) & -0.46 & $1.91\times 10^{-03}$ & 0.55 & $1.74\times 10^{-04}$ & -0.29 & $5.43\times 10^{-02}$ \\\hline
    \textbf{22.}  US Census Gini index score \cite{gini} (lower is better) & -0.44 & $3.41\times 10^{-03}$ & 0.11 & $5.12\times 10^{-01}$ & -0.5 & $6.22\times 10^{-04}$ \\\hline
    \textbf{23.}  \% with bachelor's degree or higher \cite{census} & 0.42 & $4.99\times 10^{-03}$ & -0.43 & $4.21\times 10^{-03}$ & 0.33 & $2.78\times 10^{-02}$ \\\hline
    \textbf{24.}  Avg \# poor mental health days, past 30 days \cite{america} & -0.39 & $1.05\times 10^{-02}$ & 0.1 & $5.31\times 10^{-01}$ & -0.48 & $1.23\times 10^{-03}$ \\\hline
    \textbf{25.}  Neuroticism Big Five personality trait \cite{rentfrow} & -0.37 & $1.30\times 10^{-02}$ & 0.23 & $1.35\times 10^{-01}$ & -0.37 & $1.42\times 10^{-02}$ \\\hline
    \textbf{26.}  Extraversion Big Five personality trait \cite{rentfrow} & -0.34 & $2.78\times 10^{-02}$ & 0.13 & $4.13\times 10^{-01}$ & -0.29 & $5.36\times 10^{-02}$ \\\hline
    \textbf{27.}  Farmers markets per 100,000 in pop. \cite{produce} & 0.33 & $2.88\times 10^{-02}$ & -0.01 & $9.59\times 10^{-01}$ & 0.42 & $5.41\times 10^{-03}$ \\\hline
    \textbf{28.}  Binge drinking rate \cite{america} & 0.33 & $2.88\times 10^{-02}$ & -0.12 & $4.88\times 10^{-01}$ & 0.41 & $6.23\times 10^{-03}$ \\\hline
    \textbf{29.}  Avg \# poor physical health days, past 30 days \cite{america} & -0.32 & $3.83\times 10^{-02}$ & 0.16 & $3.32\times 10^{-01}$ & -0.38 & $1.16\times 10^{-02}$ \\\hline
    \textbf{30.}  Strolling of the Heifers locavore score (lower is better) \cite{stroll} & -0.31 & $4.52\times 10^{-02}$ & -0.16 & $3.32\times 10^{-01}$ & -0.45 & $3.16\times 10^{-03}$ \\\hline
    \textbf{31.}  \% schools offering fruit/veg at celebrations \cite{produce} & 0.25 & $1.13\times 10^{-01}$ & -0.38 & $1.36\times 10^{-02}$ & 0.05 & $7.75\times 10^{-01}$ \\\hline
    \textbf{32.}  Openness Big Five personality trait \cite{rentfrow} & 0.23 & $1.30\times 10^{-01}$ & -0.42 & $5.43\times 10^{-03}$ & 0.04 & $7.95\times 10^{-01}$ \\\hline
    \textbf{33.}  \% cropland harvested for fruits/veg \cite{produce} & 0.18 & $2.58\times 10^{-01}$ & -0.53 & $2.90\times 10^{-04}$ & -0.04 & $7.95\times 10^{-01}$ \\\hline
    \textbf{34.}  Conscientiousness Big Five personality trait \cite{rentfrow} & -0.1 & $5.31\times 10^{-01}$ & 0.14 & $3.97\times 10^{-01}$ & -0.05 & $7.78\times 10^{-01}$ \\\hline
    \textbf{35.}  \% census tracts, healthy food retailer within 1/2 mile \cite{produce} & -0.06 & $7.41\times 10^{-01}$ & -0.39 & $1.09\times 10^{-02}$ & -0.24 & $1.28\times 10^{-01}$ \\\hline
    \textbf{36.}  George Mason overall freedom ranking \cite{freedom} (lower is freer) & -0.02 & $8.82\times 10^{-01}$ & -0.05 & $7.73\times 10^{-01}$ & -0.1 & $5.58\times 10^{-01}$ \\\hline
    \textbf{37.}  Agreeableness Big Five personality trait \cite{rentfrow} & 0 & $9.85\times 10^{-01}$ & 0.24 & $1.26\times 10^{-01}$ & 0.08 & $6.41\times 10^{-01}$
    \\\hline
  \end{tabular}
}{}
  \caption{
        Identical to Tab.~\ref{tab:fluxwell.corrTable} but including liquids and using a caloric difference
    rather than caloric ratio.
    Spearman correlation coefficients, $\rhospearman$, and Benjamini-Hochberg $q$-values for
    caloric input $\calin$,
    caloric output $\calout$, 
    and caloric difference $\caldiff(\alpha) = \alpha \calout + (1-\alpha)\calin$
    and
    demographic data related to food and physical activity, Big Five
    personality traits~\cite{rentfrow}, health and well-being rankings by state, and
    socioeconomic status, correlated, ordered from strongest to weakest Spearman
    correlations with caloric ratio.
    We chose $\alpha$ so that the average of $\calout$ matched the
    average of $\alpha \calin$.
  }
  \label{tab:fluxwell.supp.corrTableLIQ_DIF}
  \plosoneonly{\end{adjustwidth}}
\end{table*}

\end{document}